# The Software Complexity of Nations

Sándor Juhász[a,b,c,d], Johannes Wachs[b,c,d*], Jermain Kaminski[e], César A. Hidalgo[f,g,h]

[a] Corvinus Institute for Advanced Studies, Corvinus University of Budapest
[b] Institute for Data Analytics and Information Systems, Corvinus University of Budapest, Hungary
[c] Centre for Economic and Regional Studies, ELTE, Budapest, Hungary
[d] Complexity Science Hub, Vienna, Austria
[e] School of Business and Economics, Maastricht University, Netherlands
[f] Center for Collective Learning, IAST, Toulouse School of Economics, University of Toulouse Capitole, France
[g] Center for Collective Learning, CIAS, Corvinus University of Budapest, Hungary
[h] Alliance Manchester Business School, University of Manchester, United Kingdom
*Corresponding author: johannes.wachs@uni-corvinus.hu

**Abstract**
Despite the growing importance of the digital sector, research on economic complexity and its implications continues to rely mostly on administrative records—e.g. data on exports, patents, and employment—that have blind spots when it comes to the digital economy. In this paper we use data on the geography of programming languages used in open-source software to extend economic complexity ideas to the digital economy. We estimate a country's software economic complexity index (ECIsoftware) and show that it complements the ability of measures of complexity based on trade, patents, and research to account for international differences in GDP per capita, income inequality, and emissions. We also show that open-source software follows the principle of relatedness, meaning that a country's entries and exits in programming languages are partly explained by its current pattern of specialization. Together, these findings help extend economic complexity ideas and their policy implications to the digital economy.

**Acknowledgements**
César A. Hidalgo acknowledges funding by the European Union project 101086712-LearnData-HORIZON-WIDERA-2022-TALENTS-01 financed by European Research Executive Agency (REA) (https://cordis.europa.eu/project/id/101086712), IAST funding from the French National Research Agency (ANR) under grant ANR-17-EURE-0010 (Investissements d'Avenir program), and the European Lighthouse of AI for Sustainability [grant number 101120237-HORIZON-CL4-2022-HUMAN-02]. Sándor Juhász's work was supported by the European Union's Marie Skłodowska-Curie Postdoctoral Fellowship Program (SUPPED, grant number 101062606). Johannes Wachs acknowledges support from the Hungarian National Scientific Fund (OTKA FK 145960). Jermain Kaminski acknowledges the support of the Dutch Research Council (NWO) through the Starter Grant. The authors would like to thank Viktor Stojkoski for support in the multidimensional estimates of economic complexity and the econometric models. César A. Hidalgo is a founder of Datawheel LLC and a creator of the Observatory of Economic Complexity (oec.world).

## 1. Introduction

The study of economic complexity has predominantly relied on administrative records, such as international trade data (Hidalgo et al., 2007; Hidalgo and Hausmann, 2009), patent filings (Balland and Rigby, 2017; Kogler et al., 2013), and employment statistics (Jara-Figueroa et al., 2018; Neffke and Henning, 2013), that while valuable, struggle to capture the importance of the digital economy. This "dark matter" (Greenstein and Nagle, 2014) is important because software capabilities—which are human capital intensive—represent a mobile and transmissible source of economic complexity that is relevant for policy efforts focused on increasing the complexity of economies (Hidalgo, 2023). Yet, despite this evident need, internationally comparable estimates of software-related economic complexity remain limited.

Economic complexity refers to the structure and breadth of productive capabilities embedded or implicit in an economy's industries, products, or workforce (Hidalgo & Hausmann, 2009, Hausmann et al. 2014, Hidalgo 2021). Methodologically, its modeled using two key concepts: the *economic complexity index (ECI)* and the idea of *relatedness*.

The economic complexity index (ECI) provides a mean to estimate the combined presence of an economy's capabilities without having to define them (Hidalgo and Stojkoski, 2025). It is often used to anticipate macroeconomic outcomes, such as long-term economic growth (Hidalgo and Hausmann 2009, Domini 2022, Chavez et al. 2017, Stojkoski et al. 2023), since economies endowed with diverse capabilities can recombine them into more complex and highervalue added products (Hidalgo and Stojkoski, 2025). Relatedness asserts that regions and countries diversify into new activities when these share capabilities with those that an economy is currently specialized in (Hidalgo et al. 2007, Neffke et al. 2011, Neffke et al. 2013, Hausmann et al. 2014,

Hidalgo et al., 2018, Hidalgo, 2021, Balland et al. 2022). For instance, a country with expertise in data analytics and high-performance computing is more likely to expand into fields that build upon that foundation, such as artificial intelligence, than countries lacking these complementary specializations.

While economic complexity methods have expanded to include trade, patents, employment, and research publication data, their application to the digital sector remains limited. Software capabilities are only partially visible in these metrics and digital capabilities are insufficiently expressed in physical product data (Rahmati et al., 2021; Stojkoski et al., 2024). Code crosses borders through cloud services, downloads, and remote platforms rather than through customs, and digital firms often create local subsidiaries that obscure trade flows even further. Moreover, service trade categories remain notoriously broad (including groupings such as "computer and information services"); and patents record protectable inventions rather than the open knowledge embedded in everyday programming.

Yet, these data limitations are at odds with the growing importance of the digital economy and the role played by open-source software (OSS). IT technologies and software development are predictors of firm productivity, innovation capacity, and economic growth (Brynjolfsson and Hitt, 2003, 1998; Rahmati et al., 2021). Within this sector, OSS libraries have become essential building blocks (Eghbal, 2020), with OSS participation predicting higher entrepreneurial activity (Wright et al., 2023) and value-added productivity in ecosystems with complementary capabilities (Nagle, 2019, 2018; Rock, 2019). In the US alone, annual investment in OSS were estimated to be about $38bn in 2019 (Korkmaz et al., 2024), and government subsidies to OSS generate large returns

(Gortmaker, 2025). As it is known for complex and innovative activities (Audretsch and Feldman, 1996; Balland et al., 2020), OSS development is human capital-intensive geographically concentrated (Wachs et al., 2022), and open to international collaboration (Goldbeck, 2025). This suggests software capabilities may follow spatial patterns distinct from traditional complexity metrics.

Taken together, the growing importance of the digital economy, the key role that open-source software plays in it, and the remaining open questions about the geography of software capabilities, represent a critical gap in economic complexity research. Moreover, it remains unclear whether the "complexity" of the digital economy substitutes or complements traditional complexity metrics. In this paper, we address these gaps by exploring the question: *Do economic complexity measures based on the geography of open-source software production correlate with macroeconomic indicators like GDP per capita, inequality, and emissions, complementing complexity measures based on trade, research, and patents?*

In this study, we use data on the geographic distribution of OSS projects hosted on GitHub to generate a national-level software economic complexity index ($ECI^{software}$). Our main specification constructs $ECI^{software}$ from clusters of programming languages frequently used together in repositories. The cluster-based measure summarizes the diversity and sophistication of a country's software capabilities in a way that is comparable across countries and aligned with how developers combine technologies in practice. We then link $ECI^{software}$ to GDP per capita, inequality measured through the Gini coefficient, and $CO_2$-per-GDP from the World Bank and compare its explanatory power with complexity indices based on trade, patents, and research. Our analyses show that $ECI^{software}$ captures a digital capability dimension that while correlated with trade-, patent- and research-based complexity measures ($R^2$~0.5-0.6) adds significant explanatory power in cross-country models of GDP per capita and income inequality. In addition, we show that countries' entries and exits in programming languages follows the principle of relatedness, confirming that digital diversification mirrors path-dependence observed in physical industries.

By incorporating software into the complexity toolbox, we provide evidence that digital specialization is reshaping economic structures and creating new pathways for structural transformation. From a policy perspective, the accessibility and granularity of open-source software data offers a cost-effective and reproducible means to track and potentially enhance economic complexity research, providing policymakers a new route to design interventions focused on fostering digital capabilities. Unlike traditional development strategies focused on infrastructure and physical capital, fostering digital complexity relies more on human capital development and knowledge spillovers within software ecosystems (Apostol and Hernández-Rodríguez, 2024; Balland et al., 2022; Brynjolfsson and Saunders, 2010; Korkmaz et al., 2024), and thus, represents a new frontier for applied and fundamental work in Econ. Geogr. and economic complexity research.

## 2. Economic complexity and open-source software production

### 2.1 Complexity, relatedness and the digital sector

Economic complexity involves the use of fine-grained data on activities to capture economic structure and shifts in specialization patterns (Balland et al., 2022; Domini, 2022; Guevara et al., 2016; Hausmann et al., 2014; Hidalgo et al., 2018, 2007; Hidalgo, 2021; Hidalgo and Hausmann, 2009; Hidalgo and Stojkoski, 2025; Poncet and de Waldemar, 2015; Stojkoski et al., 2023b). These structural measures are used to explain  variation in macroeconomic outcomes, such as economic growth (Pérez-Balsalobre, 2019; Chávez et al., 2017; Domini, 2022; Hausmann et al., 2014; Hidalgo and Hausmann, 2009; Koch, 2021; Ourens, 2012; Poncet and de Waldemar, 2013; Stojkoski et al., 2016, 2023b; Weber et al., 2021), income and gender inequality (Bandeira Morais et al., 2018; Ben Saâd and Assoumou-Ella, 2019; Chu and Hoang, 2020; Hartmann et al., 2017; Lee and Vu, 2019; Sbardella et al., 2017), and emissions (Can and Gozgor, 2017; Doğan et al., 2021; Lapatinas et al., 2019; Mealy and Teytelboym, 2020; Romero and Gramkow, 2021). In the last fifteen years, these methods grew into popular indicators for international and regional development policy (Balland et al., 2022; Hidalgo, 2023, 2021) together with methods designed to explain shifts in specialization, building on the principle of relatedness (Hidalgo et al., 2018): the notion that economies diversify by entering activities that reuse some of their existing capabilities. Relatedness metrics highlight path dependencies and help predict which industries, products, research activities, or technologies are likely to grow or decline in a country, city, or region (Alabdulkareem et al., 2018; Apostol and Hernández-Rodríguez, 2024; Boschma et al., 2013; Guevara et al., 2016; Hidalgo et al., 2018, 2007; Jara-Figueroa et al., 2018; Kogler et al., 2013; Li and Neffke, 2024; Neffke et al., 2011; Neffke and Henning, 2013; Poncet and de Waldemar, 2015). Complexity metrics then provide a comparative estimate of the value of a region's specialization pattern.

But while economic complexity methods enjoy significant adoption in policy and academia, their application is still limited by the availability of fine-grained data. Like the proverbial man looking for his keys under the lamppost, economic complexity efforts thus far have focused on international trade statistics (Hidalgo et al., 2007; Hidalgo and Hausmann, 2009), manufacturing, payroll, firm registry, and employment data for industries (Chávez et al., 2017; Fritz and Manduca, 2021; Gao and Zhou, 2018; Hidalgo, 2021; Jara-Figueroa et al., 2018; Neffke et al., 2011; Neffke and Henning, 2013), data on occupations (Alabdulkareem et al., 2018; Farinha et al., 2019; Jara-Figueroa et al., 2018; Muneepeerakul et al., 2013), patents (Balland and Rigby, 2017; Kogler et al., 2013), and research papers (Chinazzi et al., 2019; Guevara et al., 2016; Stojkoski et al., 2023b). This expansion recently led to the introduction of multidimensional economic complexity (Stojkoski et al., 2023b), the notion that metrics of complexity derived from multiple datasets complement each other to explain macroeconomic outcomes (e.g. trade and patent complexity estimates explain economic growth better together than alone). But with the exception of some recent work on digital trade (Stojkoski et al., 2023a), digital infrastructure (Liang and Tan, 2024), and software components in physical products (Rahmati et al., 2021), the multidimensional expansion of economic complexity is yet to fully reach the digital sector, despite work highlighting the importance of software outside economic complexity research (Shapiro and Varian, 1999, Chattergoon and Kerr, 2022).

For instance Aum and Shin (2024) emphasize the critical role played by software in modern economies, highlighting how it substitutes labor with high elasticity. Branstetter et al. (2019) find that firms, not only technology firms, with greater software intensity measured by patenting activity achieve greater returns to R&D. These results suggest that data on software activity can predict macro level growth. Moreover, the growth of the digital economy and its integration into the offline economy is thought to reduce greenhouse emissions (Liu et al., 2023; Zhang et al., 2024). The impact of digitalization and software production on inequality outcomes is less clear, as unequal access and winner take all dynamics may compound inequality (Arthur, 1994), while growth in access to information and employment opportunities may shrink it (Tian et al., 2025).

In practice the question of how software complexity influences macroeconomic outcomes like growth, inequality and emissions, remains unclear because economic complexity research still suffers from a “digital blind-spot”. This is due to the lack of datasets that capture a detailed view of software-related activity (Balland and Rigby, 2017; Chávez et al., 2017; Guevara et al., 2016; Stojkoski et al., 2023b). This gap hinders our ability to apply the insights derived from other datasets to digital industries, making it difficult to—for instance—forecast which digital diversification efforts are more likely to succeed or estimate how software capabilities evolve and cluster over time.

There is in fact some evidence hinting to the notion that data used traditionally to study economic complexity can miss digital capabilities. Economic complexity estimates derived from trade data (Hidalgo and Hausmann, 2009) may not align well with software, which crosses borders through cloud services, downloads, or remote platforms rather than through standard customs channels (Corrado et al., 2005; Stojkoski et al., 2023a). As a result, trade data may systematically underestimate digital activity. Service trade data should be an alternative, but it is notoriously coarse, with categories such as “Computer and Information Services”, which are too broad to distinguish basic IT outsourcing from advanced software development. Moreover, software production is often carried out through subsidiaries, obscuring the real geography of capabilities. Furthermore, open-source projects and collaborative code repositories do not appear as discrete tradeable goods (Greenstein and Nagle, 2014; Korkmaz et al., 2024) since many software products are monetized via subscriptions, advertising, or freemium models, making them hard to track in conventional trade records. When it comes to employment statistics, software is also represented through coarse industry categories, such as “Software Publishing,” and coarse occupations, such us “Software developers” which provide no information about the programming languages used or the applications created by this segment of the labor force.

In short, it is difficult to describe an economy’s digital capabilities using traditional data sources. This limits our understanding of the path-dependent dynamics and sophistication of digital economies. Countries or regions that excel in certain digital fields may not show up clearly in traditional complexity data, undercutting our ability to understand related diversification in their context. More generally, we cannot tell how productive capabilities in this sector relate to important macroeconomic outcomes such as income, growth, inequality and the carbon intensity of economies. Digital or software complexity may complement or substitute classic economic complexity estimates, which are significant predictors of these outcomes. But to understand whether these are complements or substitutes, we need to test these ideas empirically.

### 2.2 Conceptualizing software complexity

Insofar we have argued that data used to commonly estimate economic complexity fails to capture information about an economy’s digital capabilities. But what data can we use to approximate capabilities implicit in the digital economy? Here, we follow a two-pronged approach, building on data on *programming languages* and *software bundles*.

Programming languages provide an unusually fine-grained and consistent trace of digital production. A language is not only a syntax but a technical paradigm formed by an ecosystem of tools, libraries, and conventions that shapes how software is built and maintained (Valverde and Solé, 2015a,b). Language adoption indicates embedded knowledge and skills: familiarity with syntax, common practices, and domain-focused applications (e.g., AI, cybersecurity, or high-performance computing).

Languages are also meaningful categories because their ecosystems exhibit strong social and market dynamics. The value of adopting a language often depends on the availability of complementary assets—libraries, frameworks, documentation, and experienced developers—so technology choices reflect local talent pools and ecosystem maturity rather than purely technical merits (Meyerovich and Rabkin, 2013). These complementarities generate switching costs: the primary barrier to adopting a new language is frequently the surrounding toolchain and library landscape rather than the syntax itself (Shrestha et al., 2022). As a result, language portfolios tend to evolve in path-dependent ways, with organizations moving to technologically proximate ecosystems (e.g., within enterprise stacks or within data science stacks) rather than jumping arbitrarily. For these reasons, programming languages can play a role in software-based comparisons of economies that is analogous to product categories or technology classes in traditional complexity measures: they are observable, comparable across places, and tied to capability accumulation.

Languages, however, are not the natural "activity unit" of software production: most modern software systems rely on bundles of languages that are used together as part of a coherent development stack (e.g., front-end web, data science, low-level systems). Treating each language as an independent activity risks fragmenting what practitioners and firms would recognize as a single capability bundle. To align the measurement unit with how software diversification is typically conceptualized—around software genres, use cases, and ecosystems rather than individual technologies—we aggregate languages into clusters based on their revealed co-use within repositories (Boudreau, 2012; Cennamo and Santaló, 2019). The key idea is that repeated co-use identifies stable bundles of complementary capabilities: languages that are frequently used together tend to be part of the same development stack, and these stacks are closer to the activities whose diversification and sophistication economic complexity methods are designed to capture. in patent-based complexity, patent classes are already higher-level, use-oriented groupings rather than the underlying set of technologies used to produce the patent. Analogously, our co-use clusters summarize software capability bundles rather than individual syntaxes, while still being grounded in observable production choices.

In the empirical analysis, we therefore treat languages as the underlying building blocks and use software bundles as the main unit of observation. We construct these clusters using a project-level dataset of all public GitHub repositories active up to 2024 and the set of programming languages used in each repository. These clusters are interpretable as capability bundles—e.g., a front-end web stack (HTML/CSS/JavaScript), a data science stack (Python/Jupyter Notebook), or low-level systems tooling (C/Assembly/Makefile)—and provide a tractable and stable basis for country-level specialization patterns. We additionally compute versions based on individual languages, theoretically defined language groupings, and GitHub topics; these are used only as robustness checks and reported in the Supplementary Information.

### 2.3 Scope and contribution

Traditional approaches to economic complexity overlook much of the software sector's intangible and rapidly evolving nature. Programming languages, in particular clusters of languages defined by complementary use, offer a way to fill this gap by reflecting embedded knowledge, illustrating specialized skills, and revealing path-dependent growth patterns.

Specifically, we address economic complexity's digital gap by using data on the country level geographic distribution of programming languages and bundles used in OSS projects to estimate economic complexity for the software sector and explore the principle of relatedness in the context of OSS. This work does not aim to introduce a new method to estimate economic complexity, but simply to apply an existing method to new data and explore the complementarity of these estimates to those derived from well-known data sources (product exports, patents, and research publications). We acknowledge that there has been considerable work exploring alternative mathematical definitions of economic complexity, such as the transformational complexity measure (Natera and Castellacci, 2021), the Log Product Diversity (Inoua, 2023), the Ability index (Bustos and Yıldırım, 2022), and the Fitness complexity (Tacchella et al., 2012). Unlike these contributions, our paper does not involve the introduction of a new mathematical definition but the application of the Hidalgo and Hausmann (2009) definition of economic complexity to open-source software data.

In the next section we present the data and methods used to calculate these indicators and then explore their ability to explain international variance in GDP per capita, income inequality, and emissions that is unaccounted for by measures of complexity based on trade, patents, and research papers. We then construct a network of related open-source software bundles to explore the principle of relatedness in the context of software.

## 3. Data and the construction of economic complexity measures

We begin by describing the data sources and methods used to construct the country–activity matrices used in the complexity analysis. A key step is that we treat programming languages as an observable building blocks of software production but aggregate them into the software bundles (a.k.a. technology stacks) used in practice. We then apply the standard economic complexity methods to this country–bundle matrix. Finally, we construct a software bundle relatedness network to test the principle of relatedness.

We use data on the geography of open-source software provided by the GitHub Innovation Graph (GHIG)[1]. GitHub is the leading platform for OSS development, with over 100 million users worldwide. The dataset presents the number of GitHub users pushing code—uploading local code from a developer's machine to an online repository—by country and programming language on a quarterly basis starting from Q1 2020 and continuing until Q4 2023. GHIG data assigns software contributions to countries based on the IP address of the developer. This data provides a more accurate measure of a location's software activity than sources relying on self-reported locations, which are known to suffer from bias (Hecht et al. 2011). After completing the basic data cleaning procedures explained in section S1 of the Supplementary information, we are left with a sample of 163 countries and 150 programming languages for the period of 2020-2023.

[1] GitHub Innovation Graph https://github.com/github/innovationgraph

To define the activity categories used in our main ECI[software] specification, we group programming languages into clusters based on their complementary use within repositories. We build these clusters from a separate project-level dataset constructed as follows. First, we identified GitHub repositories that were active in 2024 using GHArchive. Second, for each active repository we queried the GitHub GraphQL API to retrieve its set of programming languages. Repositories typically contain multiple languages; we restrict attention to the set of languages that overlap with the 150 languages retained in the GitHub Innovation Graph (GHIG) sample.

We then construct weighted language occurrence and co-occurrence counts in a way that prevents highly polyglot repositories from dominating similarity estimates. For each repository with n distinct in-scope languages, we assign each language a weight of $1/n$, so that the total language weight contributed by a repository adds to 1. For each unordered language pair within the repository, we assign a weight of $2/[n(n-1)]$, so that the total pair weight also adds to 1 for repositories with $n > 1$. Aggregating these weights across repositories yields (i) weighted marginal counts $c_l$ for each language $l$, and (ii) weighted co-occurrence counts $c_{l\,l'}$ for each pair $(l, l')$. From these counts we compute cosine similarity between languages. For languages $l$ and $l'$, cosine similarity is defined as:

$$s_{l\,l'} = \frac{c_{l\,l'}}{\sqrt{c_l}\sqrt{c_{l'}}}$$

We convert similarity to distance as: $d_{l\,l'} = 1 - s_{l\,l'}$, and apply hierarchical agglomerative clustering to this distance matrix (linkage as implemented in our code). We obtain our baseline partition by cutting the dendrogram at a distance threshold chosen to yield an interpretable set of clusters (59 in the baseline). Each programming language is assigned to exactly one software bundle or co-use cluster.

Finally, we map GHIG language-level country activity into a country–bundle matrix by summing over languages within each bundle. Let $X_{cl}$ denote the number of developers in country c pushing code in language l (from GHIG). For each cluster $k$, we define:

$$X_{ck} = \sum_{l \in k} X_{cl}$$

This country–bundle matrix $X_{ck}$ is the main input to our construction of $ECI^{software}$ below. In the Supplementary Information (S1, S3, S4), we present three alternative operationalizations of $ECI^{software}$, based on individual languages, theoretical clusters of languages derived from the computer science literature, and topics (user tags of project content).

We estimate the Economic Complexity Index (ECI) using the standard technique introduced by (Hidalgo and Hausmann, 2009). Let $X_{ck}$ be a matrix counting the number of developers with an IP in country $c$ pushing code to GitHub in software bundle $k$. We use $X_{ck}$ to derive the matrix of specialization or revealed comparative advantage $R_{ck}$ as:

$$R_{ck} = \frac{X_{ck}X}{X_c X_k},$$

where omitted indexes have been added over (e.g. $X_c = \sum_k X_{ck}$). We then binarize the matrix $R_{ck}$ to generate the matrix $M_{ck} = 1$ if $R_{ck} \geq 1$ or 0 otherwise. Finally, we let the economic complexity index of a country $c$ ($ECI_c$) and the software bundle complexity index of an activity $k$ ($PCI_k$) be defined as the stead state of the map:

$$ECI_c = \frac{1}{M_c} \sum_k M_{ck} PCI_k$$

$$PCI_k = \frac{1}{M_k} \sum_c M_{ck} ECI_c$$

As is customary, we normalize ECI and PCI values by subtracting their respective mean and dividing them by their standard deviation.

There are several interpretations of ECI. In the context of a supply side production function, it is a method to recover an economy's capabilities from a matrix of geographic specialization (Hidalgo and Stojkoski, 2025). ECI is also a spectral-clustering method that identifies whether an economy belongs to the high- or low-capability cluster, by assigning a number to each economy and to each activity that minimizes the distance between the number assigned to each economy and the numbers assigned to each activity (Bottai et al., 2024; Mealy et al., 2019; Servedio et al., 2024). That is, it provides an optimal one-factor split of the specialization matrix. From an intuitive perspective, the capability interpretation of economic complexity simply means that higher complexity economies tend to be endowed with more of the complementary factors of production needed to specialize in activities.

We compare ECI indicators derived from open-source software ($ECI^{software}$) with the multidimensional economic complexity data compiled by (Stojkoski et al., 2023b), which uses trade data from the Observatory of Economic Complexity (oec.world), patent data from the World Intellectual Property Organization's International Patent System, and research publication data from SCImago Journal & Country Rank portal. These datasets are described in detail in section S5 of the Supplementary information.

We explore the ability of $ECI^{software}$ to complement traditional economic complexity measures in explaining international variation in GDP per capita, income inequality, and emissions. All macroeconomic indicators are derived from the Databank of The World Bank. We use simple cross-sectional Ordinary Least Squares (OLS) models, based on around 90 observations, since the relatively short coverage of the GHIG data (four years) limits our analysis to controlled correlation tests.

We test the principle of relatedness following the approach introduced in the product space (Hidalgo et al., 2007), which starts from the same specialization matrix (*M*) we used to derive measures of economic complexity. Formally, we define the proximity between two software bundles *k* and *k'* as the minimum of the two conditional probabilities that a country specialized in one is also specialized in the other:

$$\phi_{kk'} = \frac{\sum_c M_{ck} M_{ck'}}{\max(M_k, M_{k'})}$$

And define the relatedness between a county *c* and a software bundle *k* as:

$$\omega_{ck} = \frac{\sum_{k'} M_{ck'} \phi_{kk'}}{\phi_k}$$

Where again, missing indices have been added over (e.g. $\phi_k = \sum_{k'} \phi_{kk'}$).

To assess whether countries are more likely to enter software bundles related to their existing portfolio of open-source software specializations, we run linear probability models with country and language-cluster fixed effects. We estimate relatedness using 2020 data and say that a country enters a software bundle if they were not specialized in that software bundle (RCA < 1) in 2020 and 2021 and then gained comparative advantage (RCA>=1) in 2022 and 2023 (e.g. $M_{ck}$={0,0,1,1} for the years 2020 to 2023). Our models predict entry as a function of relatedness and software bundle ubiquity.

## 4. Results

### 4.1. Software and economic complexity

We begin our analysis by comparing our estimate of economic complexity based on the geography of programming languages clusters ($ECI^{software}$), with published estimates of economic complexity based on physical product exports ($ECI^{trade}$), patents ($ECI^{technology}$), and research publications ($ECI^{research}$) (Stojkoski et al., 2023b).

Figure 1A compares four specialization matrices (*M*) where countries are sorted by diversity (number of products, software bundles they specialize in, etc.) and columns are sorted by ubiquity (number of countries specialized in each software bundle, product, etc.). Much like the specialization matrices for trade, patents, and research papers, the *country-software bundle* matrix exhibits a nested structure (Bustos et al., 2012; Mariani et al., 2019), meaning that low diversity economies tend to specialize in a subset of ubiquitous activities found in more diverse economies.

Figure 1B shows a map of $ECI^{software}$ based ranking of countries constructed from the country-software bundle matrix and Figure 1C compares $ECI^{software}$ with the three other ECI measures, showing that the geography of software complexity is different from that expressed in data on products, patents, and research publications. For instance, Russia (RUS), a well-known natural resource exporter with a low $ECI^{trade}$ score (0.112 on a normalized [-1,1] scale), scores much higher in $ECI^{software}$ (0.872 on a normalized [-1,1] scale). Similarly, India (IND) scores much higher in $ECI^{software}$ (0.606) than in $ECI^{research}$ (-0.633). The contrast between software and the other dimensions is highlighted by cases such as Indonesia (IDN) and Pakistan (PAK), which rank relatively high in $ECI^{software}$ (0.872 and 0.225) despite scoring much lower in the other ECI measures. Section S6 of the Supplementary information presents a table comparing the values of $ECI^{software}$, $ECI^{trade}$, $ECI^{technology}$, and $ECI^{research}$ for all countries in our sample.

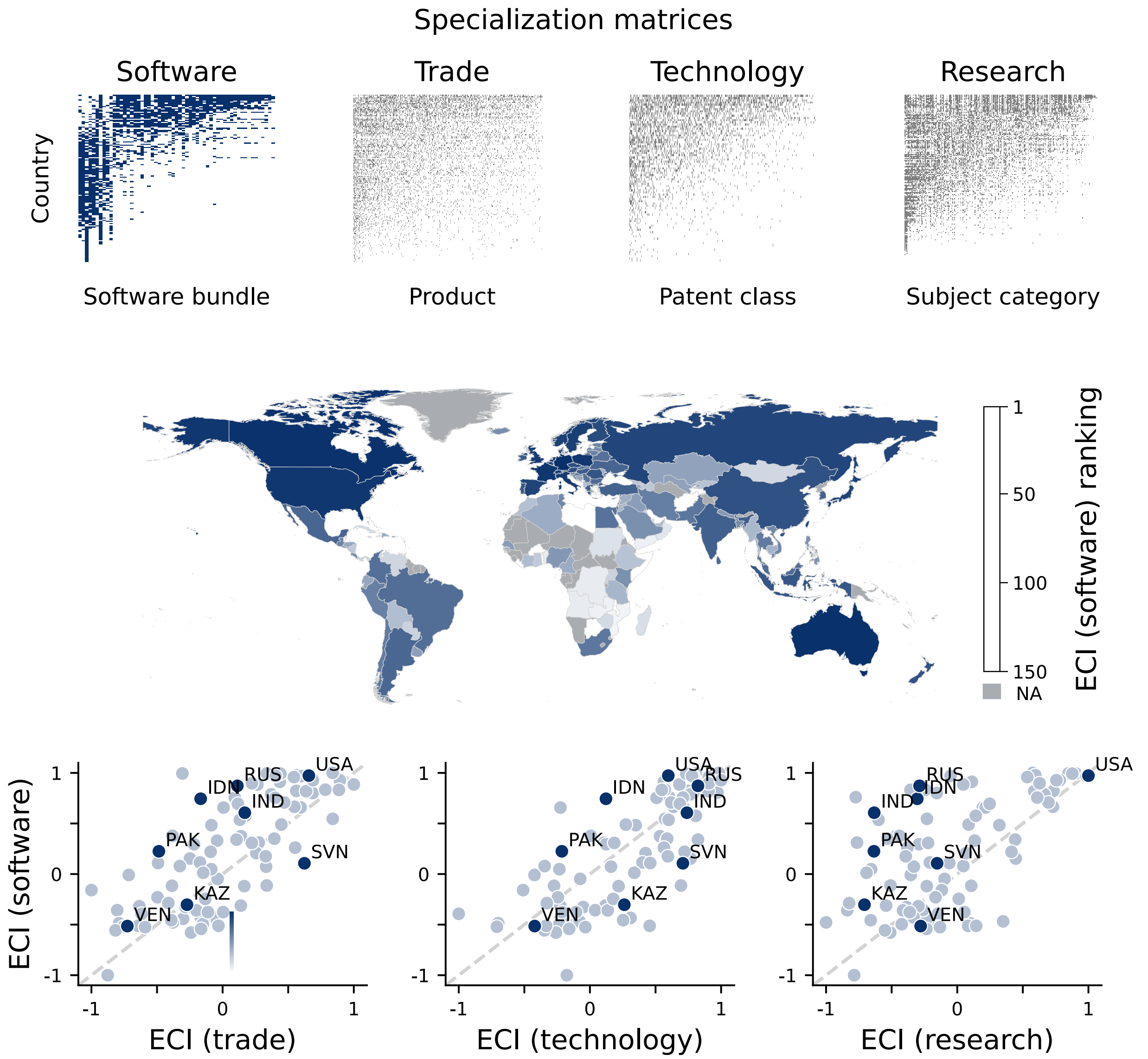

**Figure 1** **A** Specialization matrices for countries and software bundles, products, patents, and research papers. **B** Geographic distribution of software economic complexity ($ECI^{software}$). **C** Comparison between $ECI^{software}$ and $ECI^{trade}$, $ECI^{technology}$, and $ECI^{research}$ respectively ($R^2$=0.576, p-value <0.001, $R^2$=0.620, p-value <0.001 and $R^2$=0.346, p-value <0.001). For visualization purposes, ECI values are normalized to a scale of [-1, 1]. All ECI measures presented above are calculated using 2020 data only.

Next, we explore whether $ECI^{software}$ complements other measures of economic complexity in explaining international variation in GDP per capita, income inequality, and emissions.

Descriptive statistics for the key variables are presented in section S7 of the Supplementary information.

**Table 1** $ECI^{software}$ and GDP per capita (2020) in a multidimensional setting. Robust standard errors in parentheses. Significance codes: *p<0.1, **p<0.05, ***p<0.01

| | GDP per capita (log) | | | | | | | | | |
|---|---|---|---|---|---|---|---|---|---|---|
| | (1) | (2) | (3) | (4) | (5) | (6) | (7) | (8) | (9) | (10) |
| $ECI^{software}$ | 0.343*** | 0.358*** | | | | 0.180*** | 0.192*** | 0.338*** | 0.125*** | 0.169*** |
| | (0.025) | (0.026) | | | | (0.037) | (0.037) | (0.037) | (0.044) | (0.043) |
| $ECI^{trade}$ | | | 0.337*** | | | 0.222*** | | | 0.190*** | 0.177*** |
| | | | (0.028) | | | (0.037) | | | (0.046) | (0.045) |
| $ECI^{technology}$ | | | | 0.266*** | | | 0.156*** | | 0.063* | 0.051 |
| | | | | (0.021) | | | (0.029) | | (0.035) | (0.036) |
| $ECI^{research}$ | | | | | 0.140*** | | | 0.006 | 0.022 | 0.013 |
| | | | | | (0.025) | | | (0.028) | (0.026) | (0.025) |
| Population (ln) | -0.146*** | -0.150*** | -0.079*** | -0.103*** | -0.066*** | -0.117*** | -0.133*** | -0.145*** | -0.122*** | -0.120*** |
| | (0.017) | (0.017) | (0.015) | (0.019) | (0.020) | (0.014) | (0.017) | (0.019) | (0.016) | (0.016) |
| Natural resources (ln) | 0.015 | 0.018 | 0.023* | -0.018 | -0.037** | 0.034*** | 0.007 | 0.015 | 0.028** | 0.031** |
| | (0.012) | (0.013) | (0.013) | (0.012) | (0.018) | (0.012) | (0.011) | (0.012) | (0.014) | (0.014) |
| Instrument variable | No | Yes | No | No | No | No | No | No | No | Yes |
| Observations | 93 | 93 | 93 | 93 | 93 | 93 | 93 | 93 | 93 | 93 |
| $R^2$ | 0.648 | 0.647 | 0.693 | 0.654 | 0.374 | 0.753 | 0.711 | 0.648 | 0.764 | 0.762 |

Table 1 shows that the correlation between $ECI^{software}$ and GDP per capita remains strong after controlling for other estimates of economic complexity. In fact, $ECI^{software}$ works out to be as good as $ECI^{trade}$ at explaining international variations in GDP per capita in the complete model (column 8). This validates $ECI^{software}$ as a complementary indicator by showing that there is information about international variations in GDP per capita contained in $ECI^{software}$ that is not redundant with the information captured by the other ECIs. Moreover, the robustness of results across different model specifications suggests $ECI^{software}$ is a reliable and consistent predictor. We also note that in this model $ECI^{trade}$ remains statistically significant across specifications, but $ECI^{technology}$ and $ECI^{software}$ lose their significance in the full models, suggesting that the information about international variations in GDP per capita carried by them is redundant with the information available in $ECI^{software}$ and $ECI^{trade}$.

Economic complexity indicators often show patterns of spatial clustering, as illustrated in Figure 1A. Moran's I confirms spatial autocorrelation (global Moran's I=0.483, p<0.01), suggesting that countries with similar $ECI^{software}$ values are geographically proximate, deviating significantly from a random distribution (Salinas, 2021). To address potential endogeneity issues and illustrate the robustness of our results, we provide instrumental variable (IV) regressions, following the identification strategy of (Stojkoski et al., 2023b). Detailed explanation and all the related regression results can be found in section S8 of the Supplementary information. The IV regressions in models (2) and (10) of Table 1 show results comparable to our baseline estimations.

**Table 2** $ECI^{software}$ and income inequality in a multidimensional setting. ECI estimates are based on 2020 data, while the dependent variable is the average Gini coefficient between 2020 and 2022. Robust standard errors in parentheses. Significance codes: *p<0.1, **p<0.05, ***p<0.01

| | Gini coefficient | | | | | | | | | |
|---|---|---|---|---|---|---|---|---|---|---|
| | (1) | (2) | (3) | (4) | (5) | (6) | (7) | (8) | (9) | (10) |
| $ECI^{software}$ | -1.038*** | -1.054*** | | | | -0.905** | -1. 033*** | -0.981*** | -0.920** | -0.966** |
| | (0.353) | (0.413) | | | | (0.358) | (0.409) | (0.349) | (0.381) | (0.416) |
| $ECI^{trade}$ | | | -0.679** | | | -0.500* | | | -0.359 | -0.354 |
| | | | (0.289) | | | (0.275) | | | (0.293) | (0.294) |
| $ECI^{technology}$ | | | | -0.219 | | | -0.013 | | 0.061 | 0.069 |
| | | | | (0.253) | | | (0.288) | | (0.285) | (0.281) |
| $ECI^{research}$ | | | | | 0.419** | | | 0.387** | 0.332** | 0.331** |
| | | | | | (0.158) | | | (0.144) | (0.153) | (0.154) |
| GDP per capita (ln) | 0.905*** | 0.918** | 0.612* | 0.262 | -0.330 | 1.219*** | 0.914*** | 0.521 | 0.759** | 0.787** |
| | (0.350) | (0.389) | (0.322) | (0.324 | (0.249) | (0.357) | (0.350) | (0.344) | (0.343) | (0.367) |
| Population (ln) | 0.455*** | 0.460*** | 0.222** | 0.177* | 0.090 | 0.481*** | 0.456*** | 0.401*** | 0.422*** | 0.435*** |
| | (0.129) | (0.146) | (0.088) | (0.091) | (0.078) | (0.127) | (0.125) | (0.116) | (0.113) | (0.128) |
| Natural resources (ln) | 0.250** | 0.248** | 0.286** | 0.354*** | 0.400*** | 0.224* | 0.251** | 0.313*** | 0.279** | 0.274** |
| | (0.109) | (0.112) | (0.117) | (0.112) | (0.092) | (0.117) | (0.113) | (0.097) | (0.117) | (0.121) |
| Instrument variable | No | Yes | No | No | No | No | No | No | No | Yes |
| Observations | 48 | 48 | 48 | 48 | 48 | 48 | 48 | 48 | 48 | 48 |
| $R^2$ | 0.409 | 0.409 | 0.357 | 0.299 | 0.376 | 0.445 | 0.409 | 0.484 | 0.499 | 0.499 |

Next, we look at the ability of $ECI^{software}$ to explain international variations in income inequality (Table 2). Since official data on income inequality are infrequently published, and Gini coefficients vary slowly over time, we use the average Gini coefficient from the 2020–2022 period. Despite the more limited sample, we find the same negative and significant relationship between income inequality and $ECI^{software}$. In fact, $ECI^{software}$ remains strong, negative, and significant across all specifications. We also find $ECI^{research}$ remains significant, albeit with a positive coefficient.

**Table 3** $ECI^{software}$ and greenhouse gas emission intensity (2020) in a multidimensional setting. Robust standard errors in parentheses. Significance codes: *p<0.1, **p<0.05, ***p<0.01

| | Emission per GDP (log) | | | | | | | | | |
|---|---|---|---|---|---|---|---|---|---|---|
| | (1) | (2) | (3) | (4) | (5) | (6) | (7) | (8) | (9) | (10) |
| $ECI^{software}$ | -0.115*** | -0.112** | | | | -0.118*** | -0.106** | -0.079* | -0.072 | -0.059 |
| | (0.041) | (0.043) | | | | (0.043) | (0.047) | (0.044) | (0.050) | (0.052) |
| $ECI^{trade}$ | | | -0.021 | | | 0.012 | | | 0.001 | -0.001 |
| | | | (0.040) | | | (0.040) | | | (0.042) | (0.042) |
| $ECI^{technology}$ | | | | -0.052 | | | -0.016 | | -0.014 | -0.017 |
| | | | | (0.033) | | | (0.038) | | (0.039) | (0.039) |
| $ECI^{research}$ | | | | | -0.064*** | | | -0.046** | -0.046** | -0.048** |
| | | | | | (0.020) | | | (0.021) | (0.021) | (0.022) |
| GDP per capita (ln) | 0.011 | 0.009 | -0.051 | -0.020 | -0.031 | 0.004 | 0.019 | 0.013 | 0.019 | 0.016 |
| | (0.027) | (0.028) | (0.032) | (0.030) | (0.024) | (0.034) | (0.029) | (0.026) | (0.034) | (0.034) |
| Population (ln) | 0.031* | 0.030 | -0.005 | 0.006 | -0.002 | 0.030 | 0.032* | 0.024 | 0.025 | 0.022 |
| | (0.018) | (0.018) | (0.014) | (0.016) | (0.013) | (0.018) | (0.018) | (0.018) | (0.018) | (0.018) |
| Natural resources (ln) | 0.054*** | 0.055*** | 0.066*** | 0.067*** | 0.062*** | 0.056*** | 0.055*** | 0.053*** | 0.054*** | 0.055*** |
| | (0.013) | (0.014) | (0.015) | (0.012) | (0.012) | (0.014) | (0.013) | (0.013) | (0.015) | (0.015) |
| Instrument variable | No | Yes | No | No | No | No | No | No | No | Yes |
| Observations | 92 | 92 | 92 | 92 | 92 | 92 | 92 | 92 | 92 | 92 |
| $R^2$ | 0.553 | 0.553 | 0.506 | 0.521 | 0.557 | 0.553 | 0.554 | 0.576 | 0.577 | 0.577 |

Finally, we look at the intensity of greenhouse gas emissions (emissions per unit of GDP per capita) (Table 3). This is a particularly interesting outcome for $ECI^{software}$ because compared to the physical economy, software and information technologies are expected to be a less carbon-intensive way to generate GDP (Ciuriak and Ptashkina, 2020; Haberl et al., 2020; Hubacek et al., 2021; Romero and Gramkow, 2021; Stojkoski et al., 2023a; Wang and Zhang, 2021; Wiedenhofer et al., 2020).

Our results suggest that software complexity is negatively associated with emissions per unit of GDP in simpler specifications. However, in full models that account for multiple dimensions of complexity, this effect becomes statistically insignificant. This pattern indicates that $ECI^{software}$ and $ECI^{research}$ may share overlapping explanatory power. The factor (VIF) analysis (section S14) suggests some degree of collinearity between software and research complexity. While economies with high software complexity tend to have high research complexity (their individual effects on emissions seem to operate through distinct mechanisms, as evidenced by a non-significant interaction term we testes separately). One interpretation of these findings is that $ECI^{research}$ absorbs part of the explanatory power of $ECI^{software}$ in predicting emissions, since research-driven economies may be more likely to invest in low-carbon technologies and knowledge-intensive, low-emission industries.

Correlating $ECI^{software}$ with income inequality and emissions intensity allows us to test the Kuznets hypotheses. In section S9 of the Supplementary information, we present regressions including a squared term for GDP per capita. The results support the Kuznets hypothesis for income inequality, indicating an inverted U-shaped relationship, but show little evidence of such a pattern for emissions intensity.

### 4.2. Related diversification in open-source software

Having validated $ECI^{software}$ as a complementary measure of economic complexity, we now explore whether changes in the software specialization of countries is subject to the principle of relatedness: the notion that economies are more likely to enter—and less likely to exit—related activities (Autant-Bernard, 2001; Guevara et al., 2016; Hidalgo et al., 2018, 2007; Jaffe, 1986; Neffke et al., 2011; Neffke and Henning, 2013).

Table 4 present our linear probability models predicting entry events as a function of relatedness and the ubiquity of a software bundle or language cluster. We also include country and bundle fixed effects and employ clustered standard errors by country to account for within-country correlations over time, ensuring robust and reliable standard errors in our regression models. Estimations based on logit models can be found in section S10 of the Supplementary information.

**Table 4** Entry models on countries gaining revealed comparative advantage (RCA >= 1) in software bundles (2020-2023). Standard errors are clustered at the country level. Significance codes: *p<0.1, **p<0.05, ***p<0.01

| | Entry | | | | | | |
|---|---|---|---|---|---|---|---|
| | (1) | (2) | (3) | (4) | (5) | (6) | (7) |
| Relatedness density | 0.154** | 0.349** | 0.282*** | 0.429** | | 0.171** | 0.328** |
| | (0.072) | (0.133) | (0.097) | (0.174) | | (0.079) | (0.134) |
| Ubiquity | | | | | -0.006 | -0.012 | -0.012 |
| | | | | | (0.009) | (0.010) | (0.010) |
| Country FE | No | Yes | No | Yes | No | No | Yes |
| Software bundle FE | No | No | Yes | Yes | No | No | No |
| Observations | 764 | 764 | 764 | 764 | 764 | 764 | 764 |
| $R^2$ | 0.013 | 0.187 | 0.118 | 0.271 | 0.001 | 0.016 | 0.189 |

Table 4 suggests that open-source software specialization follows the principle of relatedness, with countries being more likely to specialize in software bundles that are related to those they are currently specialized in. The negative and significant effect of bundle ubiquity indicates that countries are less likely to enter common language bundles, which is reasonable since many countries already have comparative advantage in them. While relatedness in the case of OSS behaves similarly across both simpler and more complex models, its explanatory power remains limited, with a baseline $R^2$ of about 1%. We suggest a few reasons why this is still a significant finding. First, entry is a rare event: we observe 42 entrances vs 722 non-entrances. Second, the R-squared values of the models with country and language-cluster fixed effects are much higher (27%) and the estimate of the effect of relatedness on entry is about three times as large as in the baseline model (0.154 vs 0.429). Third, similar levels of explanatory power are observed in other papers testing the principle of relatedness (for example see Balland et al., 2018; and for a general overview see Li and Neffke, 2024). Interpreting the effect size also indicates the significance of relatedness as a correlate of entry. The mean of the relatedness measure in the full sample is 0.326, with a standard deviation of 0.168. Moving from the mean to one standard deviation above it is associated with a 7.2–percentage-point increase in the probability of entry, nearly double the base rate of entry of 5-6% to about 12-13%.

Figure 2 shows the network of related software bundles following the visualization approach of (Hidalgo et al., 2007). Figure 2A highlights a few example software bundles, with labels listing all programming languages within each. We then focus on the entry and exit patterns of three countries on Figure 2B. In each case, entries occur into bundles that are adjacent to existing specializations, while exits tend to occur out of more weakly connected bundles.

Figure 2B highlights contrasting dynamics in countries' software capability portfolios, measured as entries and exits in revealed comparative advantage (RCA) across software bundles. China exhibits multiple entries, consistent with an expanding and diversifying software profile: it is increasingly likely to develop comparative advantage in additional capability bundles, suggesting active broadening of its OSS specializations. Great Britain shows comparatively few transitions, indicating a more stable specialization structure over the period—its portfolio appears to evolve gradually, with limited reallocation across bundles. Russia, in contrast, displays several exits, consistent with a contraction or relative weakening of specialization in a set of capability bundles, likely related to large scale emigration of software developers in the wake of the 2022 invasion of Ukraine (Wachs, 2023).

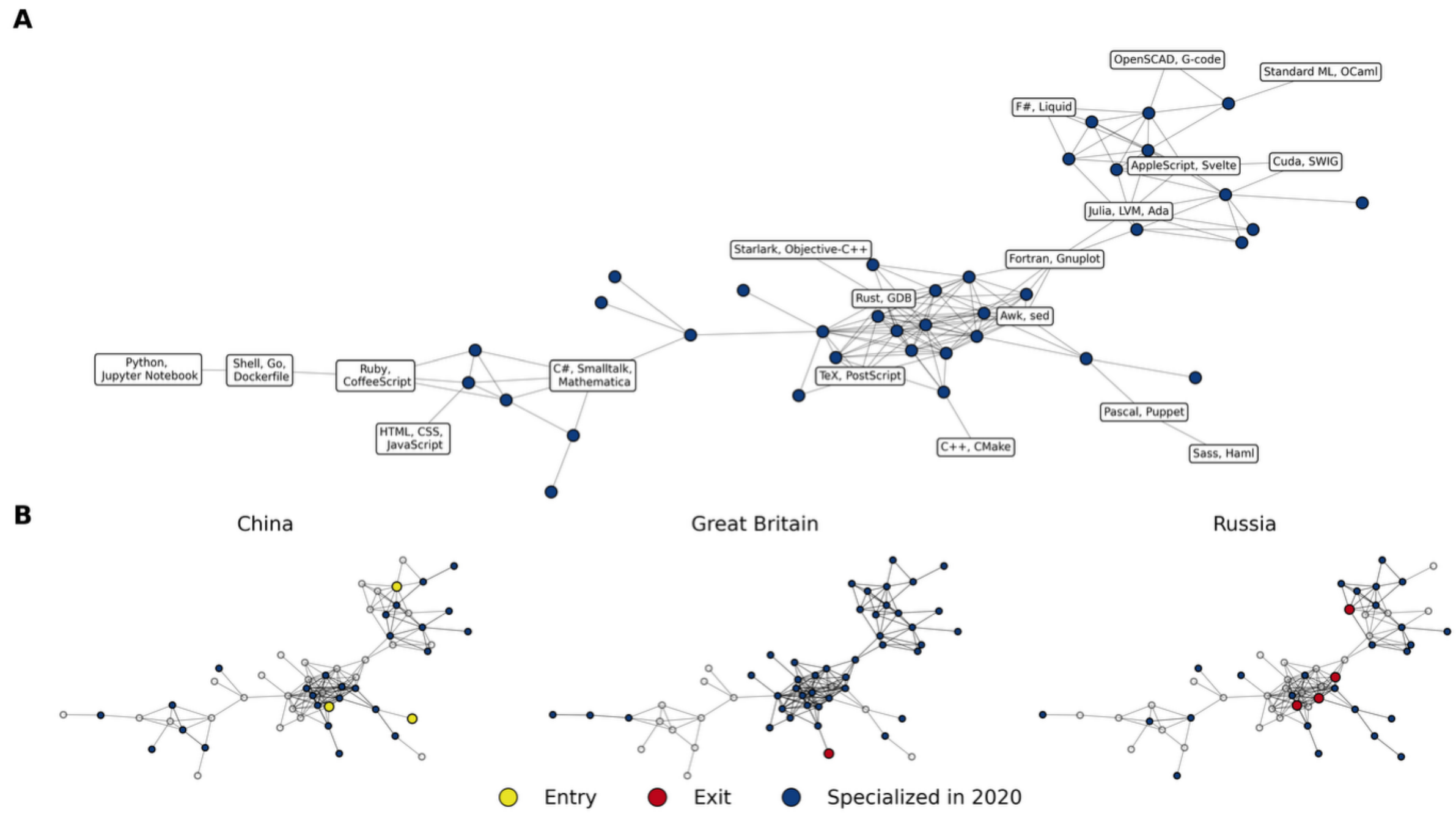

**Figure 2 (A)** Network representation of software bundle relatedness. **(B)** Changes in revealed comparative advantage (RCA) in programming languages clusters (2020-2023) in China, Great Britain, and Russia. Dark blue nodes indicate specialization in 2020-2021 (RCA >=1), while yellow nodes indicate subsequent (2022-2023) specialization in software bundles, and red nodes indicate exits. Countries are more likely to specialize in new software bundles adjacent to their previous specializations.

We then explore the principle of relatedness in the context of exits (Table 5). We consider exits as countries that were specialized in a software bundle (RCA >= 1) in 2020 and 2021 and later lost their comparative advantage (RCA < 1) in 2022 and 2023 (e.g. $M_{cl}$={1,1,0,0} for the years going from 2020 to 2023). The negative and significant effect of relatedness across both simpler and more complex models indicates that countries are less likely to lose their advantage in software bundles that are related to those they currently specialize in. Again, the effects of relatedness are overall mild ($R^2$<3% on the baseline model) but are robust to the inclusion of country and bundle fixed-effects, showing that they go beyond what can explained based on the statistic characteristics of a country or bundle.

**Table 5** Exit models on countries losing revealed comparative advantage (RCA < 1) in software bundles (2020-2023). Standard errors are clustered at the country level. Significance codes: *p<0.1, **p<0.05, ***p<0.01

| | Exit | | | | | | |
|---|---|---|---|---|---|---|---|
| | (1) | (2) | (3) | (4) | (5) | (6) | (7) |
| Relatedness density | -0.160*** | -0.405*** | -0.190*** | -0.285** | | -0.223*** | -0.348*** |
| | (0.033) | (0.105) | (0.044) | (0.116) | | (0.043) | (0.099) |
| Ubiquity | | | | | -0.006 | -0.027*** | -0.018** |
| | | | | | (0.006) | (0.008) | (0.009) |
| Country FE | No | Yes | No | Yes | No | No | Yes |
| Software bundle FE | No | No | Yes | Yes | No | No | No |
| Observations | 1544 | 1544 | 1544 | 1544 | 1544 | 1544 | 1544 |
| $R^2$ | 0.023 | 0.185 | 0.116 | 0.257 | 0.001 | 0.035 | 0.187 |

### 4.3. Robustness checks and alternative approaches

We verify the consistency of our findings through multiple alternative specifications and modeling strategies. First, we confirm that the main results hold when varying RCA thresholds or applying Tobit regressions to account for the nature of the dependent variables (see section S10 and S11 in the Supplementary information). We also verify that restricting the sample to countries with fully available macroeconomic data does not alter the significance or direction of our coefficients, indicating that sample selection does not drive our conclusions (see section S13 in the Supplementary information). Further, to address potential statistical concerns, we check for multicollinearity through VIF analyses and remove mathematical dependencies from key variables, ensuring that the variables used are valid and adequately capture different dimensions of complexity (see section S14 in the Supplementary information for more details).

Second, we go back to our alternative definitions of $ECI^{software}$ to show that our conclusions hold when we define software complexity on different basis, either by grouping languages into theoretical clusters (e.g., web-oriented or system-level languages; see S3) or by using a measure based on topics (S4), or simply by consider languages themselves (S1). We find that even when we change the unit of observation to topics, $ECI^{software}$ remains positively correlated with GDP per capita and negatively correlated with income inequality.

Our findings on the relationship between $ECI^{software}$ and macroeconomic indicators are based on cross-sectional regressions. In section S15 of the Supplementary information, we replicate GDP growth models in the style of Hidalgo and Hausmann (2009). However, this is not recommended due to the limited time span of available data (2020–2023), since measures of complexity are structural measures that are connected to long term growth (so we should not expect significance in short time periods dominated by other dynamics, such as the covid bounce-back in this case). As expected, we find that neither $ECI^{software}$ nor $ECI^{trade}$ significantly predicts GDP growth. Structural measures such as $ECI^{software}$ tend to be stable over time, whereas short-term growth outcomes are more volatile. Supporting this, we find that $ECI^{software}$ remains highly stable across years, with correlations exceeding 0.92 (see section S16 of the Supplementary information), suggesting its predictive power may become more apparent over longer time horizons. Additionally, we provide an extensive explanation of our instrumental variable approach, including extended models and tests in section S8 of the Supplementary information. However, testing for potential endogeneity using instruments for other complexity measures—or between complexity measures themselves, such as $ECI^{software}$ and $ECI^{technology}$—was beyond the scope of this paper. Together, these tests demonstrate that our main results are stable and robust, even when we account for alternative definitions, model specifications, and potential sources of bias.

## 5. Discussion

Here we expanded the study of economic complexity to include the software sector by leveraging recently published data on the geography of open-source software (OSS). By relying on the IP addresses of the developers contributing to OSS projects, instead of on self-reported locations (which can suffer from reporting bias (Hecht et al., 2011)), we were able to construct estimates of the geographic distribution of open-source software language knowledge for 100+ programming languages and use them to create internationally comparable estimates of economic complexity for the software sector and to study OSS's diffusion in the context of the principle of relatedness. Our study provides a cross-country measure of software economic complexity and demonstrates it complements well-established ECI metrics based on trade, patents and research.

Building on prior studies linking software specialization to broader skill formation and productivity gains (Brynjolfsson and Hitt, 2003; Nagle, 2019, 2018; Wright et al., 2023), our results indicate that countries with higher software-based economic complexity may be better equipped to generate inclusive growth—thereby reducing inequality. This aligns with research showing that knowledge-intensive economies can create wider opportunities for high-skilled labor, mitigating income disparities (Hartmann et al., 2017). Although not consistently significant across all models, the observed negative association between software complexity and emissions aligns directionally with prior evidence that digitally driven economies may reduce their reliance on resource-intensive activities (Haberl et al., 2020; Stojkoski et al., 2024). These points suggest that software complexity could serve as a policy-relevant indicator for steering economies toward less environmentally taxing activities. In sum, our study contributes to the literature by offering both an empirical measure of software capabilities and an interpretation, consistent with earlier scholarship, of how these capabilities might shape pathways of inclusive and sustainable growth.

We also found that $ECI^{software}$ complements other measures of economic complexity when explaining macro-outcomes. One plausible interpretation of this complementarity is that the overlap between these different activities is not exhaustive, and hence, the differences among them are informative. Patent data includes many non-software activities, such as patents in biotech or the life sciences. Similarly, research publication data also includes many non-software related sectors, such as publications in history or philosophy. Also, open-source software data may provide some additional granularity that might not be available in the other data sources. For example, OSS data involves hundreds of unique languages, which provide a resolution over the software sector that is larger than the one captured in research publication data. The idea that correlated measures of complexity can prove to be complementary is at the core of the idea of multidimensional complexity (Stojkoski et al. 2023), which is based on the idea that information on the geography of different activities (products, patents, papers, software, etc.) captures different levels of detail making them mutually reinforcing. In simple terms, they fill each other's "gaps."

But what can we make of these findings? First, that economic complexity measures derived from OSS production do indeed correlate significantly with GDP, inequality, and emissions suggests that software complexity can suggest productive diversification directions. The literature on economic development is rife with work advising economies to diversify towards more complex economic activities (Balland et al., 2018; Hausmann et al., 2014; Hidalgo, 2023). High economic complexity activities are associated with better wages and may face less competition in international markets than the production of more ubiquitous commodities. The question that remains is whether this advice can translate to software. We argue that many of the unique aspects of software make it especially attractive for specific kinds of diversification strategies.

Unlike physical products, software relies less on immobile factors, such as large manufacturing or processing plants and natural resources. At the same time, software outputs are highly tradable (OECD, 2023; Stojkoski et al., 2024) and digital products are known to be—on average—of relatively high complexity compared to physical products (Stojkoski et al., 2024). Further, transformer models on platforms like Hugging Face make deep learning accessible with pre-trained models that require significantly fewer resources (Wolf et al., 2019). This means that software provides new opportunities for structural upgrading that are less reliant on physical factors of production and more reliant on efforts to attract human capital. Combined with our finding that diversification in software follows the principle of relatedness, policymakers should seek to attract experts in complex software technologies most related to current areas of strength. Future research could explore how AI-driven productivity gains might alter the rate at which regions diversify into more sophisticated software niches—and whether that facilitates or hinders upward movement in the digital value chain.

While our study suggests how to estimate, validate, and use measures of economic complexity based on software, it is also subject to several important limitations that may affect the interpretation of our results. First, because our data exclusively captures open-source software (OSS) activity on GitHub, we may underestimate important proprietary or closed-source capabilities—and overlook OSS activity on other platforms. This can lead us to systematically undervalue software complexity in certain economies (for instance, where non-GitHub or closed-source development is predominant). Even OSS projects hosted outside of GitHub are also different on average, for example they are more likely to be academic (Trujillo et al., 2022). Moreover, our assumption that GitHub-based OSS specialization reflects broader digital skills—while supported by research on OSS's role in innovation— may still introduces measurement error. Ultimately, some countries may possess stronger software capabilities than our metrics reveal, which could influence the strength of the observed correlations with macroeconomic outcomes.

Second, applying product-complexity methods to programming languages poses conceptual challenges. We treat languages as distinct units of analysis, a choice which offers clear interpretability but simplifies the complex relationships between them. For instance, languages may relate through complementary usage (e.g., HTML and CSS) rather than hierarchical supply chains, meaning the "distance" between them may not perfectly map onto traditional complexity notions. We explored alternative specifications, such as considering individual languages or theoretical clusters instead of bundles as the basis for the ECI calculation in our robustness checks (see Supplementary Information). While these aggregations largely confirm our results, we retain the software bundle approach in our main analysis for its robustness. Ultimately, path-dependent software diversification may follow different patterns than those in manufacturing, and more granular data (e.g., at the project or framework level) will be valuable for future work.

Nevertheless, despite these limitations, our work represents a valuable step towards extending economic complexity analysis to the digital realm, offering insights into the geographic distribution of software capabilities and their potential impact on macroeconomic outcomes. Software complexity is a significant complement to trade, research, and technology complexity measures because it covers a specific and important class of capabilities; this is demonstrated by its ability to extend the predictive power of models of key macro-outcomes including growth, inequality, and emission intensity. As the digital economy continues to evolve, further research integrating diverse data sources will be crucial. Understanding how emerging technologies, particularly in artificial intelligence (Daniotti et al., 2025; Del Rio-Chanona et al., 2024), may alter the nature of software capabilities and pathways for diversification remains a key challenge for the future.

# Supplementary Information

### S1 GitHub data on programming languages and data preparation

We leverage the open access datasets by GitHub's Innovation Graph (GHIG). Software economic complexity is calculated from the *languages.csv* table that presents the number of GitHub users pushing code by country and programming language on a quarterly basis. The country of users is estimated using the IP address of each contributor. While not perfect, IP geolocation is a considerably more reliable indicator of the geography of software production than self-reported location, which can contain fictional information (e.g. Narnia, Hogwarts, etc.). The raw data captures the activity of tens of millions of developers from 164 countries in 379 languages between 2020 January and 2023 December on a quarterly basis (with regular updates). As an initial data cleaning, we excluded data formats and markup languages such as *yaml*, *json*, *text*, *svg, Markdown* and *xml* following Del Rio-Chanona et al. (2024).

To focus on the most relevant language, we limit our exercise to the top 150 languages with the most contributors on average across the 2020-2023 period. We aggregate the quarterly data to yearly observations by considering the average number of developers in each country, language combination.

Table S1.0 compares the software bundle measurement of $ECI^{software}$ for the year 2021 with alternative implementations using, respectively: individual language use data, theoretical clusters of languages derived from computer science concepts (described in S3), and topics, which are tags that users give to describe their projects (S4). Despite important differences in the definition of each of these indices, we find relatively strong correlations among the four of them. This suggests the overall robustness of data derived from programming language use for describing capabilities.

**Table S1.0** Correlation of $ECI^{software}$ based on programming languages, clusters of programming languages and topics. Correlations are based on 125 countries with available topic data for 2021

| | $ECI^{software}$ (languages) | $ECI^{software}$ (theoretical clusters) | $ECI^{software}$ (co-occurrence clusters) | $ECI^{software}$ (topics) |
|---|---|---|---|---|
| $ECI^{software}$ (languages) | 1 | 0.982 | 0.973 | 0.839 |
| $ECI^{software}$ (theoretical clusters) | 0.982 | 1 | 0.968 | 0.823 |
| $ECI^{software}$ (co-occurrence clusters) | 0.973 | 0.968 | 1 | 0.817 |
| $ECI^{software}$ (topics) | 0.839 | 0.823 | 0.8174 | 1 |

Below we present our main results based on individual programming languages as the unit of observation for the $ECI^{software}$ calculations, instead of software bundles; the results remain consistent and become slightly stronger.

**Table S1.1** $ECI^{software}$ based on programming languages and GDP per capita (2020) in a multidimensional setting. Robust standard errors in parentheses. Significance codes: *p<0.1, **p<0.05, ***p<0.01

| | GDP per capita (log) | | | | | | | | | |
|---|---|---|---|---|---|---|---|---|---|---|
| | (1) | (2) | (3) | (4) | (5) | (6) | (7) | (8) | (9) | (10) |
| $ECI^{software}$ | 0.331*** | 0.331*** | | | | 0.191*** | 0.208*** | 0.337*** | 0.155*** | 0.170*** |
| | (0.022) | (0.022) | | | | (0.028) | (0.031) | (0.034) | (0.032) | (0.034) |
| $ECI^{trade}$ | | | 0.337*** | | | 0.205*** | | | 0.171*** | 0.166*** |
| | | | (0.028) | | | (0.034) | | | (0.041) | (0.041) |
| $ECI^{technology}$ | | | | 0.266*** | | | 0.139*** | | 0.058* | 0.054 |
| | | | | (0.021) | | | (0.028) | | (0.034) | (0.035) |
| $ECI^{research}$ | | | | | 0.140*** | | | -0.009 | 0.009 | 0.006 |
| | | | | | (0.025) | | | (0.026) | (0.024) | (0.024) |
| Population (ln) | -0.141*** | -0.141*** | -0.079*** | -0.103*** | -0.066*** | -0.118*** | -0.133*** | -0.143*** | -0.116*** | -0.119*** |
| | (0.016) | (0.016) | (0.015) | (0.019) | (0.020) | (0.014) | (0.016) | (0.018) | (0.015) | (0.016) |
| Natural resources (ln) | 0.021 | 0.021 | 0.023* | -0.018 | -0.037** | 0.037*** | 0.013 | 0.021 | 0.031** | 0.033** |
| | (0.013) | (0.013) | (0.013) | (0.012) | (0.018) | (0.013) | (0.013) | (0.013) | (0.014) | (0.014) |
| Instrument variable | No | Yes | No | No | No | No | No | No | No | Yes |
| Observations | 93 | 93 | 93 | 93 | 93 | 93 | 93 | 93 | 93 | 93 |
| $R^2$ | 0.683 | 0.683 | 0.693 | 0.654 | 0.374 | 0.771 | 0.736 | 0.683 | 0.779 | 0.778 |

**Table S1.2** ECI$^{software}$ based on programming languages and income inequality in a multidimensional setting. ECI estimates are based on 2020 data, while the dependent variable is the average Gini coefficient between 2020 and 2022. Robust standard errors in parentheses. Significance codes: *p<0.1, **p<0.05, ***p<0.01

| | Gini coefficient | | | | | | | | | |
|---|---|---|---|---|---|---|---|---|---|---|
| | (1) | (2) | (3) | (4) | (5) | (6) | (7) | (8) | (9) | (10) |
| ECI$^{software}$ | -1.259*** | -1.282*** | | | | -1.133*** | -1.281*** | -1.182*** | -1.148*** | -1.289*** |
| | (0.340) | (0.351) | | | | (0.344) | (0.403) | (0.312) | (0.352) | (0.369) |
| ECI$^{trade}$ | | | -0.679** | | | -0.431 | | | -0.325 | -0.309 |
| | | | (0.289) | | | (0.276) | | | (0.265) | (0. 264) |
| ECI$^{technology}$ | | | | -0.219 | | | 0.050 | | 0.108 | 0.133 |
| | | | | (0.253) | | | (0.289) | | (0.274) | (0.279) |
| ECI$^{research}$ | | | | | 0.419** | | | 0.363*** | 0.312** | 0.307** |
| | | | | | (0.158) | | | (0.132) | (0.137) | (0.139) |
| GDP per capita (ln) | 1.279*** | 1.303*** | 0.612* | 0.262 | -0.330 | 1.522*** | 1.250*** | 0.888** | 1.062*** | 1.167*** |
| | (0.383) | (0.404) | (0.322) | (0.324) | (0.249) | (0.364) | (0.353) | (0.367) | (0.342) | (0.361) |
| Population (ln) | 0.580*** | 0.588*** | 0.222** | 0.177* | 0.090 | 0.594*** | 0.577*** | 0.519*** | 0.532*** | 0.575*** |
| | (0.144) | (0.149) | (0.088) | (0.091) | (0.078) | (0.142) | (0.139) | (0.129) | (0.127) | (0.136) |
| Natural resources (ln) | 0.183* | 0.180* | 0.286** | 0.354*** | 0.400*** | 0.166 | 0.176 | 0.247** | 0.211* | 0.192 |
| | (0.104) | (0.103) | (0.117) | (0.112) | (0.092) | (0.115) | (0.107) | (0.100) | (0.116) | (0.121) |
| Instrument variable | No | Yes | No | No | No | No | No | No | No | Yes |
| Observations | 48 | 48 | 48 | 48 | 48 | 48 | 48 | 48 | 48 | 48 |
| $R^2$ | 0.468 | 0.468 | 0.357 | 0.299 | 0.376 | 0.494 | 0.469 | 0.534 | 0.546 | 0.544 |

**Table S1.3** ECI$^{software}$ based on programming languages and greenhouse gas emission intensity (2020) in a multidimensional setting. Robust standard errors in parentheses. Significance codes: *p<0.1, **p<0.05, ***p<0.01

| | Emission per GDP (log) | | | | | | | | | |
|---|---|---|---|---|---|---|---|---|---|---|
| | (1) | (2) | (3) | (4) | (5) | (6) | (7) | (8) | (9) | (10) |
| ECI$^{software}$ | -0.100** | -0.090** | | | | -0.101** | -0.088** | -0.061 | -0.051 | -0.026 |
| | (0.040) | (0.043) | | | | (0.042) | (0.044) | (0.043) | (0.048) | (0.053) |
| ECI$^{trade}$ | | | -0.021 | | | 0.005 | | | -0.002 | -0.006 |
| | | | (0.040) | | | (0.040) | | | (0.044) | (0.043) |
| ECI$^{technology}$ | | | | -0.052 | | | -0.025 | | -0.021 | -0.026 |
| | | | | (0.033) | | | (0.036) | | (0.038) | (0.038) |
| ECI$^{research}$ | | | | | -0.064*** | | | -0.049** | -0.049** | -0.054** |
| | | | | | (0.020) | | | (0.021) | (0.022) | (0.023) |
| GDP per capita (ln) | 0.010 | 0.002 | -0.051 | -0.020 | -0.031 | 0.007 | 0.023 | 0.007 | 0.019 | 0.011 |
| | (0.029) | (0.030) | (0.032) | (0.030) | (0.024) | (0.035) | (0.030) | (0.027) | (0.034) | (0.035) |
| Population (ln) | 0.027 | 0.023 | -0.005 | 0.006 | -0.002 | 0.026 | 0.029 | 0.018 | 0.020 | 0.014 |
| | (0.018) | (0.019) | (0.014) | (0.016) | (0.013) | (0.019) | (0.019) | (0.018) | (0.018) | (0.019) |
| Natural resources (ln) | 0.055*** | 0.056*** | 0.066*** | 0.067*** | 0.062*** | 0.055*** | 0.055*** | 0.055*** | 0.055*** | 0.057*** |
| | (0.014) | (0.015) | (0.015) | (0.012) | (0.012) | (0.015) | (0.014) | (0.014) | (0.015) | (0.016) |
| Instrument variable | No | Yes | No | No | No | No | No | No | No | Yes |
| Observations | 92 | 92 | 92 | 92 | 92 | 92 | 92 | 92 | 92 | 92 |
| $R^2$ | 0.543 | 0.543 | 0.506 | 0.521 | 0.557 | 0.543 | 0.547 | 0.569 | 0.571 | 0.569 |

**Table S1.4** Entry models on countries gaining revealed comparative advantage (RCA >= 1) in programming languages (2020-2023). Standard errors are clustered at the country level. Significance codes: *p<0.1, **p<0.05, ***p<0.01

| | Entry | | | | | | |
|---|---|---|---|---|---|---|---|
| | (1) | (2) | (3) | (4) | (5) | (6) | (7) |
| Relatedness density | 0.207*** | 0.262* | 0.384*** | 0.321** | | 0.241*** | 0.218* |
| | (0.064) | (0.144) | (0.081) | (0.135) | | (0.069) | (0.113) |
| Ubiquity | | | | | -0.026*** | -0.034*** | -0.048*** |
| | | | | | (0.009) | (0.009) | (0.008) |
| Country FE | No | Yes | No | Yes | No | No | Yes |
| Language FE | No | No | Yes | Yes | No | No | No |
| Observations | 1584 | 1584 | 1584 | 1584 | 1584 | 1584 | 1584 |
| $R^2$ | 0.021 | 0.095 | 0.188 | 0.277 | 0.011 | 0.038 | 0.121 |

**Table S1.5** Exit models on countries losing revealed comparative advantage (RCA < 1) in programming languages (2020-2023). Standard errors are clustered at the country level. Significance codes: *p<0.1, **p<0.05, ***p<0.01

| | Exit | | | | | | |
|---|---|---|---|---|---|---|---|
| | (1) | (2) | (3) | (4) | (5) | (6) | (7) |
| Relatedness density | -0.088*** | -0.257*** | -0.072*** | -0.308*** | | -0.112*** | -0.270*** |
| | (0.021) | (0.062) | (0.026) | (0.115) | | (0.025) | (0.073) |
| Ubiquity | | | | | -0.002 | -0.012* | 0.003 |
| | | | | | (0.006) | (0.006) | (0.010) |
| Country FE | No | Yes | No | Yes | No | No | Yes |
| Language FE | No | No | Yes | Yes | No | No | No |
| Observations | 2978 | 2978 | 2978 | 2978 | 2978 | 2978 | 2978 |
| $R^2$ | 0.009 | 0.101 | 0.097 | 0.181 | 0.000 | 0.011 | 0.101 |

## S2 GitHub project data and programming language clustering

Our main cluster of languages is based on their co-use within software projects. Here we recapitulate the clustering procedure and describe the result. In particular, we collected a novel dataset of over 30 million GitHub projects active in 2024 and the languages used in each project. We count the frequency of co-occurrence of languages across all projects. We calculate the cosine similarity of two languages as follows:

$$\text{cosine_sim}(l_1, l_2) = \frac{\text{pair_counts}(l_1, l_2)}{\sqrt{\text{language_counts}[l_1]} \times \sqrt{\text{language_counts}[l_2]}}$$

We again carry out a hierarchical clustering analysis, using Ward's distance and a cut value of 1. We obtain 59 clusters.

**Table S2.1** Clusters (Cl.) of programming languages based on co-occurrence in GitHub projects

| Cl. | Languages | Cl. | Languages |
|---|---|---|---|
| 1 | CSS, HTML, JavaScript | 31 | Brainfuck, Forth |
| 2 | C++, CMake | 32 | Fortran, Gnuplot |
| 3 | Assembly, C, Makefile | 33 | Awk, sed |
| 4 | HLSL, ShaderLab | 34 | DIGITAL Command Language, M4, Roff |
| 5 | Dart, Kotlin, Objective-C, Swift | 35 | Meson, SmPL |
| 6 | OCaml, Standard ML | 36 | Elixir, Erlang |
| 7 | Stata, SystemVerilog, Tcl, VHDL, Verilog | 37 | D, DTrace |
| 8 | Blade, Hack, PHP | 38 | Pug, Stylus |
| 9 | M, MATLAB | 39 | GDB, Logos, Rust |
| 10 | Jupyter Notebook, Python | 40 | Objective-C++, Starlark |
| 11 | Dockerfile, Go, Shell | 41 | GLSL, NSIS, Processing |
| 12 | HCL, Smarty | 42 | AMPL, Lua, Scheme |
| 13 | GAP, GDScript | 43 | Clojure, Emacs Lisp |
| 14 | Lex, Yacc | 44 | Common Lisp, Prolog |
| 15 | PLSQL, PLpgSQL, SQLPL, TSQL | 45 | Scala, XSLT |
| 16 | Batchfile, PowerShell | 46 | ANTLR, Thrift |
| 17 | ASP.NET, Visual Basic .NET | 47 | VBA, VBScript |
| 18 | C#, Mathematica, Smalltalk | 48 | Apex, OpenEdge ABL |
| 19 | Less, SCSS, TypeScript, Vue | 49 | Scilab, UnrealScript |
| 20 | QML, QMake | 50 | Haml, Sass |
| 21 | CoffeeScript, Ruby | 51 | Cuda, SWIG |
| 22 | Pascal, Puppet | 52 | Ada, Julia, LLVM |
| 23 | Pawn, SourcePawn | 53 | AutoHotkey, Inno Setup |
| 24 | Perl, Raku, XS | 54 | Handlebars, Solidity |
| 25 | FreeMarker, Gherkin, Groovy, Java | 55 | AppleScript, Nim, Svelte |
| 26 | PostScript, TeX | 56 | F#, Liquid |
| 27 | R, Rebol | 57 | NASL, Twig |
| 28 | Haskell, Nix | 58 | Elm, RobotFramework |
| 29 | Vim Script, Vim Snippet | 59 | ActionScript, Mako, PureBasic |
| 30 | G-code, OpenSCAD | | |

## S3 Theoretical clusters of programming languages and ECI software

Here we describe an alternative approach to clustering languages: first by a theoretical taxonomy of languages derived from their design properties, and second by their co-occurrence within a large scale dataset of software projects, suggesting that they are used together. We carry out these clustering exercises to show that our results are robust to alternative conceptualization of capabilities in software. Specifically, we aggregate national activity in individual languages to the cluster level and recalculate the software ECI measure.

Using this clustering approach, the 150 languages were grouped into 38 different clusters. The ECI values we derived using the countries, clusters and contributors matrix are very correlated (0.983) to the original, programming language based ECI values. The Tables below illustrate that ECI software based on theoretical clusters is similarly correlated to GDP per capita and Gini coefficient, while it does not have a significant relationship with greenhouse gas emissions. Additionally, we tested the entry and exit models and found that exit models show similar results, while relatedness density based on theoretical clusters of languages has no significant relationship with the few observed entries.

**Table S3.1** Theoretical clusters (Cl.) of programming languages

| Cl. | Languages | Cl. | Languages |
|---|---|---|---|
| 1 | Haml, Handlebars, Liquid, Smarty, Twig, Vue | 21 | F#, Scala |
| 2 | Blade, FreeMarker, Mako, Pug, QML, Svelte | 22 | ASP.NET, Apex, Visual Basic .NET |
| 3 | Makefile, Meson, Nix | 23 | C#, D, Dart, Java, Kotlin |
| 4 | CMake, Puppet, QMake | 24 | C, Cuda |
| 5 | DTrace, GDB | 25 | C++, Objective-C, Objective-C++ |
| 6 | Gherkin, SWIG, SmPL | 26 | Pascal |
| 7 | Inno Setup, NSIS, Vim Snippet | 27 | Solidity, Swift |
| 8 | ANTLR, Lex, Thrift, Yacc | 28 | Ada, Fortran, Go, Rust |
| 9 | Brainfuck, HCL | 29 | DIGITAL Command Language, Tcl |
| 10 | Dockerfile, Jupyter Notebook, M4 | 30 | Awk, Batchfile, Rebol, Shell, Vim Script, sed |
| | | | CoffeeScript, JavaScript, Julia, Lua, MATLAB, |
| | | | PHP, Perl, PowerShell, Python, R, Raku, Ruby, |
| 11 | Assembly, PLpgSQL, TSQL | 31 | Smalltalk |
| 12 | PLSQL, Processing | 32 | TypeScript |
| 13 | Forth, GLSL, VHDL | 33 | AppleScript, AutoHotkey |
| 14 | HLSL, ShaderLab, Verilog | 34 | GDScript, Mathematica, VBScript |
| 15 | Pawn, PureBasic, SourcePawn | 35 | Prolog |
| 16 | CSS, HTML, Less, SCSS, Sass, Stylus, TeX | 36 | Haskell, OCaml, Standard ML |
| 17 | GAP, Gnuplot, NASL, Starlark | 37 | Clojure, Common Lisp, Emacs Lisp, Scheme |
| 18 | G-code, M, Roff, XSLT | 38 | Elixir, Erlang |
| 19 | LLVM, OpenSCAD, PostScript, XS | | |
| 20 | Groovy, Hack | | |

**Table S3.2** Correlation of ECI software based on theoretical language clusters with macroeconomic indicators

| | GDP per capita (log) | | Gini coefficient | | Emission per GDP (log) | |
|---|---|---|---|---|---|---|
| | (1) | (2) | (3) | (4) | (5) | (6) |
| $ECI^{software}$ (theoretical clusters) | 0.321*** | 0.145*** | -1.160*** | -1.010*** | -0.104*** | -0.061 |
| | (0.023) | (0.036) | (0.306) | (0.347) | (0.038) | (0.048) |
| $ECI^{trade}$ | | 0.177*** | | -0.355 | | -0.001 |
| | | (0.044) | | (0.288) | | (0.043) |
| $ECI^{technology}$ | | 0.053 | | 0.122 | | -0.016 |
| | | (0.034) | | (0.285) | | (0.040) |
| $ECI^{research}$ | | 0.015 | | 0.263* | | -0.048** |
| | | (0.025) | | (0.140) | | (0.022) |
| GDP per capita (ln) | | | 1.230*** | 1.030*** | 0.015 | 0.023 |
| | | | (0.362) | (0.349) | (0.027) | (0.033) |
| Population (ln) | -0.134*** | -0.112*** | 0.505*** | 0.454*** | 0.027 | 0.021 |
| | (0.015) | (0.015) | (0.124) | (0.108) | (0.017) | (0.017) |
| Natural resources (ln) | 0.018 | 0.030** | 0.223** | 0.251** | 0.055*** | 0.054*** |
| | (0.013) | (0.014) | (0.101) | (0.116) | (0.013) | (0.015) |
| Observations | 93 | 93 | 48 | 48 | 92 | 92 |
| $R^2$ | 0.683 | 0.776 | 0.461 | 0.525 | 0.548 | 0.575 |

**Table S3.3** Entry and exit models on countries gaining and losing revealed comparative advantage (RCA) in theoretical clusters of programming languages (2020-2023)

| | Entry | | | Exit | | |
|---|---|---|---|---|---|---|
| | (1) | (2) | (3) | (4) | (5) | (6) |
| Relatedness density (clusters) | 0.013 | 0.028 | 0.081 | -0.196*** | -0.268*** | -0.342*** |
| | (0.064) | (0.064) | (0.086) | (0.041) | (0.051) | (0.088) |
| Ubiquity (clusters) | | -0.014* | -0.014 | | -0.034*** | -0.025* |
| | | (0.008) | (0.009) | | (0.008) | (0.013) |
| Country FE | No | No | Yes | No | No | Yes |
| Observations | 689 | 689 | 689 | 1,166 | 1,166 | 1,166 |
| $R^2$ | 5.7e-5 | 0.003 | 0.138 | 0.023 | 0.039 | 0.206 |

## S4 Topics of repositories and ECI software

To supplement our main results using contributions in programming languages to GitHub, we use an alternative dataset from GitHub's Innovation Graph (GHIG) on the most popular project topics within an economy. Precisely, we use the *topics.csv* table that gives the total count of unique developers making at least one git push to a repository with a given topic on a quarterly basis. The raw data captures the activity of tens of millions of developers from 131 countries in 1337 topics between 2020 January and 2023 December (with regular updates). To be comparable to our exercise using programming language, we excluded data topics related to formats and markup languages such as *yaml*, *json*, *text*, *svg, Markdown* and *xml* following Del Rio-Chanona et al. (2024) and focus on the top 200 topics with the most contributors on average across the 2020-2023 period. We aggregate the quarterly data to yearly observations by considering the average number of developers in each country, topic combination. We also exclude topics with less than 200 contributors across the world and remove the EU as a "country." Below, we present our main results using $ECI^{software}$ calculated from topic contributions.

Table S4.1 presents regressions based on 2020 data. The data on topics for this year—our selected period for the main analyses—is limited and only available for 55 countries. While our main findings hold for GDP per capita and income inequality, the number of observations is low. In Table S4.2, we repeat the analysis using 2021 data, which includes topic information for a larger set of countries. The results are consistent with our main findings, however, emissions data is not available for 2021, preventing us from reproducing those specific results. Tables S4.3 and S4.4 show the correlations between all four $ECI^{software}$ measures. They indicate that the topic-based $ECI^{software}$ differs somewhat from the language-cluster-based measures, though in 2021, all measures are highly correlated. Finally, Tables S4.5 and S4.6 compare the $ECI^{software}$ measures in terms of their correlations with GDP per capita and the Gini coefficient in 2021. These tables suggest that the programming-language-based $ECI^{software}$ is the most promising approach, offering higher explanatory power ($R^2$) and more significant coefficients.

**Table S4.1** Correlation of ECI software based on topics on GitHub with macroeconomic indicators (2020)

| | GDP per capita (log) | | Gini coefficient | | Emission per GDP (log) | |
|---|---|---|---|---|---|---|
| | (1) | (2) | (3) | (4) | (5) | (6) |
| $ECI^{software}$ (topics) | 0.215*** | 0.073** | -0.430 | -0.612*** | -0.007 | 0.021 |
| | (0.035) | (0.030) | (0.266) | (0.206) | (0.038) | (0.036) |
| $ECI^{trade}$ | | 0.125*** | | -0.524 | | -0.004 |
| | | (0.039) | | (0.321) | | (0.047) |
| $ECI^{technology}$ | | 0.086* | | 0.250 | | -0.047 |
| | | (0.051) | | (0.278) | | (0.037) |
| $ECI^{research}$ | | 0.061** | | 0.460** | | -0.071*** |
| | | (0.026) | | (0.178) | | (0.024) |
| GDP per capita (ln) | | | 0.276 | 0.228 | -0.016 | 0.083 |
| | | | (0.273) | (0.324) | (0.047) | (0.075) |
| Population (ln) | -0.218*** | -0.114*** | 0.517** | 0.619*** | 0.038 | 0.025 |
| | (0.026) | (0.023) | (0.234) | (0.183) | (0.035) | (0.032) |
| Natural resources (ln) | -0.019 | -0.011 | 0.178 | 0.264** | 0.063*** | 0.072*** |
| | (0.013) | (0.011) | (0.130) | (0.113) | (0.018) | (0.016) |
| Observations | 51 | 51 | 33 | 33 | 50 | 50 |
| R2 | 0.718 | 0.876 | 0.403 | 0.590 | 0.562 | 0.637 |

**Table S4.2** Correlation of ECI software based on topics of repositories with macroeconomic indicators (2021)

| | GDP per capita (log) | | Gini coefficient | |
|---|---|---|---|---|
| | (1) | (2) | (3) | (4) |
| $ECI^{software}$ (topics) | 0.295*** | 0.131*** | -0.527* | -0.766*** |
| | (0.036) | (0.039) | (0.262) | (0.229) |
| $ECI^{trade}$ | | 0.197*** | | -0.701*** |
| | | (0.040) | | (0.243) |
| $ECI^{technology}$ | | 0.082* | | 0.037 |
| | | (0.047) | | (0.234) |
| $ECI^{research}$ | | 0.026 | | 0.346* |
| | | (0.033) | | (0.178) |
| GDP per capita (ln) | | | 0.258 | 0.503 |
| | | | (0.221) | (0.314) |
| Population (ln) | -0.207*** | -0.152*** | 0.384** | 0.544*** |
| | (0.026) | (0.023) | (0.164) | (0.139) |
| Natural resources (ln) | 0.008 | 0.038** | 0.283** | 0.206** |
| | (0.013) | (0.016) | (0.106) | (0.102) |
| Observations | 86 | 86 | 46 | 46 |
| R2 | 0.545 | 0.773 | 0.370 | 0.515 |

**Table S4.3** Correlation of ECI software based on programming languages, clusters of programming languages and topics. Correlations are based only on 53 countries with available topic data for 2020

| | **$ECI^{software}$ (languages)** | **$ECI^{software}$ (theoretical clusters)** | **$ECI^{software}$ (co-occurrence clusters)** | **$ECI^{software}$ (topics)** |
|---|---|---|---|---|
| $ECI^{software}$ (languages) | 1 | 0.983 | 0.970 | 0.533 |
| $ECI^{software}$ (theoretical clusters) | 0.983 | 1 | 0.974 | 0.406 |
| $ECI^{software}$ (co-occurrence clusters) | 0.970 | 0.974 | 1 | 0.465 |
| $ECI^{software}$ (topics) | 0.533 | 0.406 | 0.465 | 1 |

**Table S4.4** Correlation of ECI software based on programming languages, clusters of programming languages and topics. Correlations are based on 125 countries with available topic data for 2021

| | **$ECI^{software}$ (languages)** | **$ECI^{software}$ (theoretical clusters)** | **$ECI^{software}$ (co-occurrence clusters)** | **$ECI^{software}$ (topics)** |
|---|---|---|---|---|
| $ECI^{software}$ (languages) | 1 | 0.982 | 0.973 | 0.839 |
| $ECI^{software}$ (theoretical clusters) | 0.982 | 1 | 0.968 | 0.823 |
| $ECI^{software}$ (co-occurrence clusters) | 0.973 | 0.968 | 1 | 0.817 |
| $ECI^{software}$ (topics) | 0.839 | 0.823 | 0.817 | 1 |

**Table S4.5** Regressions of different $ECI^{software}$ measures and GDP per capita (2021)

| | **GDP per capita (log)** | | | | |
|---|---|---|---|---|---|
| | (1) | (2) | (3) | (4) | (5) |
| $ECI^{software}$ (languages) | 0.403*** | | | | 0.110 |
| | (0.024) | | | | (0.119) |
| $ECI^{software}$ (theoretical clusters) | | 0.381*** | | | 0.013 |
| | | (0.023) | | | (0.147) |
| $ECI^{software}$ (co-occurrence clusters) | | | 0.413*** | | 0.218* |
| | | | (0.024) | | (0.110) |
| $ECI^{software}$ (topics) | | | | 0.385*** | 0.102* |
| | | | | (0.033) | (0.054) |
| Population (ln) | -0.145*** | -0.141*** | -0.152*** | -0.232*** | -0.180*** |
| | (0.017) | (0.018) | (0.020) | (0.025) | (0.028) |
| Natural resources (ln) | 0.027** | 0.016 | 0.019 | 0.013 | 0.028** |
| | (0.013) | (0.013) | (0.013) | (0.013) | (0.014) |
| Observations | 111 | 111 | 111 | 111 | 111 |
| R2 | 0.708 | 0.698 | 0.705 | 0.597 | 0.728 |

**Table S4.6** Regressions of different $ECI^{software}$ measures and income inequality (2021)

| | Gini coefficient | | | | |
|---|---|---|---|---|---|
| | (1) | (2) | (3) | (4) | (5) |
| $ECI^{software}$ (languages) | -1.148*** | | | | -1.325** |
| | (0.296) | | | | (0.644) |
| $ECI^{software}$ (theoretical clusters) | | -0.852*** | | | 0.241 |
| | | (0.292) | | | (0.707) |
| $ECI^{software}$ (co-occurrence clusters) | | | -0.785** | | 0.138 |
| | | | (0.326) | | (0.722) |
| $ECI^{software}$ (topics) | | | | -0.625** | -0.225 |
| | | | | (0.255) | (0.326) |
| GDP per capita (ln) | 1.071*** | 0.803** | 0.646* | 0.306 | 1.010*** |
| | (0.312) | (0.341) | (0.324) | (0.219) | (0.328) |
| Population (ln) | 0.469*** | 0.364*** | 0.329** | 0.450*** | 0.537*** |
| | (0.122) | (0.115) | (0.128) | (0.155) | (0.149) |
| Natural resources (ln) | 0.152 | 0.210** | 0.226** | 0.233** | 0.144 |
| | (0.098) | (0.095) | (0.100) | (0.095) | (0.096) |
| Observations | 111 | 111 | 111 | 111 | 111 |
| R2 | 0.421 | 0.357 | 0.345 | 0.343 | 0.433 |

## S5 Data preparation to compare economic complexity measures

We compare the economic complexity of open-source software production ($ECI^{software}$) with three other metrics of economic complexity constructed by : (1) trade complexity ($ECI^{trade}$) based on product export data from the Observatory of Economic Complexity[2], (2) technology complexity ($ECI^{technology}$) based on patent applications data from World Intellectual Property Organization's International Patent System, and (3) research complexity ($ECI^{research}$) based on published research documents data from SCImago Journal & Country Rank portal[3]. The alternative ECI indicators are constructed in the similar fashion as $ECI^{software}$ and are available for cross validation[4].

We restrict the analysis to countries with a population of more than one million, total exports of more than 1 billion USD, and at least 4 patents. In order to refine the data on research publications, we focus on countries with at least 100 publications per year in research areas where at least 30 articles are published per year. Values for country, research area combinations where fewer than 3 articles were published per year were replaced by 0 to reduce noise. Where a country, research area combination did not receive 100 citations on average in the 2017-2020 period, the value was replaced with 0.

[2] Observatory of Economic Complexity (OEC) https://oec.world
[3] SCImago Journal & Country Rank (SJR) https://www.scimagojr.com/aboutus.php
[4] https://doi.org/10.7910/DVN/K4MEFW

We connect the different versions of ECI to socio-economic indicators of countries. The economic performance of countries is measured through GDP per capita (2020) from the CEPII Gravity database (Conte et al., 2022). The income inequality and emission indicators are taken from the online data repository of the World Bank[5]. Due to the uneven data coverage, we use the average Gini coefficient of countries for the period 2010-2023. The emission intensity indicators are from 2019.

[5] World Bank https://data.worldbank.org/indicator/

## S6 Comparison of different economic complexity values

**Table S6.1** ECI values for all countries (2020) in our sample

| Ranking | Country | ECI $^{software}$ | ECI $^{trade}$ | ECI $^{technology}$ | ECI $^{research}$ | Ranking | Country | ECI $^{software}$ | ECI $^{trade}$ | ECI $^{technology}$ | ECI $^{research}$ |
|---|---|---|---|---|---|---|---|---|---|---|---|
| 1 | DEU | 1.739 | 1.895 | 1.514 | 1.507 | 51 | THA | 0.391 | 0.901 | 0.698 | -0.531 |
| 2 | AUS | 1.730 | -0.334 | 1.146 | 2.080 | 52 | CHL | 0.355 | -0.223 | 1.062 | 1.234 |
| 3 | CAN | 1.729 | 0.919 | 1.015 | 2.197 | 53 | IRN | 0.291 | -0.074 | 0.292 | -0.144 |
| 4 | NLD | 1.727 | 1.121 | 0.993 | 2.142 | 54 | PER | 0.282 | -0.696 | 0.416 | 0.168 |
| 5 | FRA | 1.702 | 1.363 | 1.079 | 1.548 | 55 | SVN | 0.278 | 1.476 | 0.939 | -0.028 |
| 6 | USA | 1.695 | 1.542 | 0.705 | 2.401 | 56 | GTM | 0.233 | -0.373 | -1.276 | 0.394 |
| 7 | POL | 1.691 | 1.049 | 1.084 | 0.189 | 57 | LTU | 0.224 | 0.908 | -0.212 | -0.401 |
| 8 | GBR | 1.687 | 1.435 | 1.107 | 2.370 | 58 | TUN | 0.183 | 0.093 | -1.039 | -1.086 |
| 9 | ITA | 1.672 | 1.321 | 1.354 | 1.419 | 59 | VNM | 0.125 | -0.025 | 0.161 | -1.160 |
| 10 | SWE | 1.620 | 1.602 | 1.551 | 1.888 | 60 | BGD | 0.090 | -1.130 | -1.438 | -0.450 |
| 11 | CHE | 1.620 | 2.003 | 1.336 | 1.939 | 61 | CRI | 0.026 | 0.189 | -0.706 | 0.092 |
| 12 | HKG | 1.595 | 1.111 | 0.634 | 0.531 | 62 | SAU | -0.081 | 0.917 | 0.909 | -0.775 |
| 13 | NOR | 1.571 | 0.698 | 1.354 | 1.617 | 63 | KEN | -0.086 | -0.489 | -1.125 | 0.520 |
| 14 | JPN | 1.552 | 2.209 | 0.883 | 0.393 | 64 | PHL | -0.091 | 0.584 | -0.091 | -0.193 |
| 15 | ESP | 1.552 | 0.779 | 1.206 | 1.591 | 65 | NGA | -0.156 | -1.684 | -1.621 | 0.047 |
| 16 | RUS | 1.530 | 0.481 | 0.481 | -0.309 | 66 | SLV | -0.247 | -0.136 | - | - |
| 17 | SGP | 1.468 | 1.787 | 0.648 | -0.219 | 67 | SEN | -0.272 | -0.704 | -1.063 | -0.053 |
| 18 | TWN | 1.464 | 1.989 | 0.601 | -0.456 | 68 | IRQ | -0.290 | -0.696 | - | -1.294 |
| 19 | BEL | 1.448 | 1.356 | 1.023 | 1.839 | 69 | URY | -0.297 | 0.004 | -0.176 | 0.320 |
| 20 | FIN | 1.444 | 1.502 | 1.349 | 1.532 | 70 | UZB | -0.365 | -0.542 | -1.240 | -1.439 |
| 21 | AUT | 1.419 | 1.543 | 1.494 | 1.558 | 71 | KAZ | -0.392 | -0.266 | 0.001 | -1.194 |
| 22 | CZE | 1.414 | 1.599 | 1.105 | -0.032 | 72 | BIH | -0.406 | 0.533 | -0.301 | -0.846 |
| 23 | DNK | 1.393 | 0.983 | 1.058 | 1.694 | 73 | ECU | -0.416 | -0.973 | -1.022 | -0.327 |
| 24 | CHN | 1.346 | 0.994 | 0.719 | -1.334 | 74 | ARM | -0.429 | -0.288 | -0.657 | -0.517 |
| 25 | NZL | 1.340 | 0.443 | 0.941 | 1.579 | 75 | HND | -0.430 | -0.602 | - | -0.341 |
| 26 | ROU | 1.335 | 1.043 | 0.517 | -0.350 | 76 | DOM | -0.477 | -0.154 | -1.012 | -0.253 |
| 27 | IDN | 1.321 | -0.063 | -0.293 | -0.346 | 77 | DZA | -0.480 | -1.301 | -0.467 | -1.470 |
| 28 | ISR | 1.261 | 1.178 | 0.752 | 1.759 | 78 | CMR | -0.480 | -1.164 | - | -0.200 |
| 29 | PRT | 1.240 | 0.490 | 0.890 | 0.816 | 79 | MDA | -0.483 | -0.126 | -0.265 | -0.575 |
| 30 | IRL | 1.192 | 1.328 | 0.791 | 1.832 | 80 | SYR | -0.492 | - | - | -1.846 |
| 31 | HUN | 1.181 | 1.420 | 0.946 | 0.752 | 81 | LBN | -0.511 | 0.271 | -0.772 | 0.410 |
| 32 | GRC | 1.179 | 0.275 | -1.022 | 0.706 | 82 | MKD | -0.512 | 0.045 | -0.995 | -0.466 |
| 33 | IND | 1.095 | 0.592 | 1.004 | -1.037 | 83 | KHM | -0.539 | -0.941 | -2.651 | -0.017 |
| 34 | TUR | 1.046 | 0.602 | 1.147 | 0.594 | 84 | TZA | -0.558 | -0.641 | - | 0.365 |
| 35 | KOR | 0.997 | 1.897 | 0.653 | -0.191 | 85 | MMR | -0.597 | -1.129 | - | -0.908 |
| 36 | UKR | 0.981 | 0.518 | 0.710 | -0.967 | 86 | JOR | -0.597 | -0.061 | -0.976 | -0.406 |
| 37 | MEX | 0.904 | 1.135 | 0.025 | 0.478 | 87 | ARE | -0.605 | 0.158 | 0.101 | -0.402 |
| 38 | ARG | 0.894 | 0.096 | 0.183 | 0.971 | 88 | CIV | -0.613 | -1.022 | - | -0.448 |
| 39 | LKA | 0.722 | -0.482 | -0.536 | -0.632 | 89 | BOL | -0.614 | -1.018 | - | -0.410 |
| 40 | BGR | 0.715 | 0.535 | 0.573 | -0.685 | 90 | ALB | -0.619 | -0.324 | -1.022 | -0.18 |
| 41 | MYS | 0.676 | 1.030 | 0.688 | -0.822 | 91 | MAR | -0.642 | -0.499 | -0.018 | -1.093 |
| 42 | BRA | 0.661 | 0.469 | 1.181 | 1.226 | 92 | KGZ | -0.645 | -0.232 | - | 0.017 |
| 43 | COL | 0.644 | 0.182 | 0.673 | 0.576 | 93 | ETH | -0.650 | -0.881 | - | 0.301 |
| 44 | BLR | 0.613 | 0.799 | -0.323 | -1.313 | 94 | GEO | -0.661 | -0.022 | -0.716 | 1.032 |
| 45 | SRB | 0.607 | 0.696 | -0.160 | -0.172 | 95 | NIC | -0.677 | -1.065 | - | - |
| 46 | EGY | 0.586 | -0.162 | -0.291 | -0.320 | 96 | AZE | -0.677 | -0.477 | -0.760 | -1.810 |
| 47 | SVK | 0.531 | 1.339 | 0.635 | -0.503 | 97 | GHA | -0.692 | -1.274 | -2.019 | 0.506 |
| 48 | PAK | 0.472 | -0.683 | -1.000 | -1.041 | 98 | KWT | -0.707 | -0.032 | -1.042 | -0.595 |
| 49 | ZAF | 0.464 | 0.085 | 0.966 | 1.167 | 99 | PAN | -0.733 | 0.201 | 0.407 | 0.471 |
| 50 | HRV | 0.442 | 0.763 | 0.341 | 0.435 | 100 | UGA | -0.733 | -0.989 | -1.251 | 0.600 |

**Table S6.2** ECI values for all countries (2020) in our sample

| Ranking | Country | ECI software | ECI trade | ECI technology | ECI research |
|---|---|---|---|---|---|
| 101 | PRY | -0.733 | -0.431 | - | 0.868 |
| 102 | RWA | -0.733 | - | - | -0.004 |
| 103 | VEN | -0.737 | -1.151 | -1.435 | -0.292 |
| 104 | MNG | -0.748 | -1.210 | -2.040 | -0.017 |
| 105 | ZWE | -0.754 | -0.888 | -0.624 | 0.018 |
| 106 | JAM | -0.754 | -0.404 | - | 0.624 |
| 107 | CUB | -0.767 | - | -2.182 | -0.29 |
| 108 | MDG | -0.78 | -1.210 | - | -0.241 |
| 109 | QAT | -0.782 | -0.057 | -0.883 | -0.2 |
| 110 | SDN | -0.806 | -1.327 | -1.279 | -0.864 |
| 111 | OMN | -0.842 | -0.206 | -1.095 | -0.78 |
| 112 | COD | -0.896 | -1.387 | - | -0.315 |
| 113 | BEN | -0.896 | - | - | -0.22 |
| 114 | AGO | -0.896 | -1.412 | - | |
| 115 | ZMB | -0.995 | -0.698 | - | 0.122 |
| 116 | YEM | -0.995 | -1.215 | - | -1.541 |
| 117 | MOZ | -1.114 | -1.189 | - | -0.004 |
| 118 | BFA | -1.531 | -1.712 | - | -0.147 |
| 119 | BWA | -1.531 | -0.575 | - | -0.942 |
| 120 | LAO | -1.531 | -0.967 | - | -0.379 |
| 121 | LBR | -1.531 | - | - | - |
| 122 | LBY | -1.531 | -1.442 | -0.920 | -1.359 |
| 123 | TJK | -1.531 | - | - | -1.410 |
| 124 | MWI | -1.531 | - | - | 0.333 |
| 125 | TGO | -1.531 | -0.857 | - | -0.999 |
| 126 | AFG | -1.531 | -1.200 | - | -0.558 |

## S7 Descriptive statistics on the key variables of our regressions

**Table S7.1** Descriptive statistics for the variables used in the regressions on $ECI^{software}$ and macroeconomic indicators

| Variable | Mean | Std. dev. | Min | Max |
|---|---|---|---|---|
| $ECI^{software}$ | 0.471 | 0.892 | -1.531 | 1.739 |
| $ECI^{trade}$ | 0.344 | 0.903 | -1.684 | 2.209 |
| $ECI^{technology}$ | 0.083 | 0.986 | -2.652 | 1.551 |
| $ECI^{research}$ | 0.207 | 1.043 | -1.810 | 2.401 |
| GDP per capita | 29,869 | 21,954 | 2,532 | 101,612 |
| Gini coefficient | 0.361 | 0.073 | 0.250 | 0.632 |
| Emission per GDP | 0.0000003 | 0.0000002 | 0.00000007 | 0.000001 |
| Population | 72,383,712 | 208,391,222 | 1,856,124 | 1,411,100,000 |
| Natural resources | 3.467 | 5.684 | 0.0002 | 29.285 |

**Table S7.2** Descriptive statistics for our key variables on entry and exit models

| Entry | Observations | Avg. relatedness density |
|---|---|---|
| 1 | 42 | 0.405 |
| 0 | 722 | 0.321 |
| **Exit** | | |
| 1 | 76 | 0.431 |
| 0 | 1468 | 0.574 |

## S8 Instrumental variables approach for assessing the impact of software on GDP, inequality and emissions

To address potential endogeneity issues and to further validate our results, we take an instrumental variables (IV) approach proposed by (Stojkoski et al., 2023b) in which we instrument the $ECI^{software}$ values of a country with the average $ECI^{software}$ values of the three most similar non-neighboring countries (countries with similar specialization patterns but no common land or maritime borders). The idea is that there might be factors that are either local (e.g., culture, geography) or relevant only to certain dependent variables (e.g., country-specific social policies to mitigate inequalities) that could drive both complexity and other macroeconomic outcomes.

To decouple local factors and conditions from our complexity estimates, we identify the three non-neighboring countries with the most similar specialization pattern (using minimum conditional probability) and take the average of their $ECI^{software}$ values. Table S8.1 illustrates the first- and second-stage IV regressions.

For each model, two diagnostic tests were performed to assess the strength of the instrumental variables. First, the Weak Instruments Test (Kleibergen & Paap, 2006) confirms the instrument's strength, as the Kleibergen-Paap rk Wald F-statistics are well above the critical threshold ($F > 10$). Second, the Durbin-Wu-Hausman test (Hausman, 1978; Wu, 1974) examines whether $ECI^{software}$ is endogenous. The Durbin-Wu-Hausman p-values suggest significant endogeneity concerns for the GDP models in both the baseline ($p = 0.036$) and full specification ($p = 0.012$), while the Gini and emissions models show no significant endogeneity. Despite the endogeneity indicated in the GDP models, the IV estimates closely match the OLS coefficients in direction and size. We include the IV specification as a robustness check in Models (2) and (10) of Tables 1, 2, and 3 in the main text.

**Table S8.1** Instrument strength, endogeneity, and overidentification tests in 2SLS regressions. Robust standard errors in parentheses. Significance codes: *p<0.1, **p<0.05, ***p<0.01

| | **Baseline Model** | | | **Full Model** | | |
|---|---|---|---|---|---|---|
| | (1) | (2) | (3) | (4) | (5) | (6) |
| Dependent variable | GDP per capita (log) | Gini coefficient | Emission per GDP (log) | GDP per capita (log) | Gini coefficient | Emission per GDP (log) |
| Endogenous Variable (Instrumented) | $ECI^{software}$ | $ECI^{software}$ | $ECI^{software}$ | $ECI^{software}$ | $ECI^{software}$ | $ECI^{software}$ |
| $ECI^{software}$ (similar, non-neighbors) | 0.824*** | -1.016*** | -0.112*** | 0.388*** | -0.931** | -0.059 |
| | (0.058) | (0.377) | (0.042) | (0.095) | (0.366) | (0.050) |
| Population (ln) | -0.345*** | 0.443*** | 0.030* | -0.276*** | 0.419*** | 0.022 |
| | (0.039) | (0.133) | (0.018) | (0.036) | (0.113) | (0.017) |
| Natural resources (ln) | 0.042 | 0.239** | 0.055*** | 0.072** | 0.265** | 0.055*** |
| | (0.029) | (0.102) | (0.013) | (0.031) | (0.106) | (0.014) |
| GDP per capita (ln) | | 0.885** | 0.009 | | 0.758** | 0.016 |
| | | (0.355) | (0.027) | | (0.323) | (0.033) |
| $ECI^{trade}$ | | | | 0.407*** | -0.340 | -0.001 |
| | | | | (0.100) | (0.258) | (0.040) |
| $ECI^{technology}$ | | | | 0.116 | 0.066 | -0.017 |
| | | | | (0.079) | (0.246) | (0.037) |
| $ECI^{research}$ | | | | 0.030 | 0.321** | -0.049** |
| | | | | (0.056) | (0.136) | (0.021) |
| Observations | 93 | 48 | 92 | 93 | 48 | 92 |
| R-squared | 0.647 | 0.409 | 0.553 | 0.762 | 0.499 | 0.577 |
| Kleibergen-Paap (KP) LM | 28.755 | 15.991 | 33.293 | 30.058 | 14.002 | 31.050 |
| KP Underidentification p-value | 0.001 | 0.001 | 0.001 | 0.001 | 0.001 | 0.001 |
| KP rk Wald F-stat | 1955.84 | 283.78 | 983.71 | 537.44 | 258.94 | 547.599 |
| Durbin-Wu-Hausman Chi2 | 4.403 | 0.010 | 0.052 | 6.299 | 0.102 | 0.493 |
| Durbin-Wu-Hausman p-value | 0.036 | 0.922 | 0.819 | 0.012 | 0.750 | 0.483 |

*Notes:* Except for $ECI^{software}$ (similar, non-neighbors), the instrumental variable reported in first stage, the coefficients shown in the table represent the second-stage results of the regression. The reported diagnostic statistics refer to the first stage of the 2SLS estimation. The Underidentification Test (Kleibergen-Paap LM) examines whether the instrument is correlated with the endogenous regressor; rejecting the null suggests the instrument is valid. The corresponding p-value indicates whether this rejection is statistically significant ($p < 0.05$ suggests a strong instrument). The Weak Instrument Test (Kleibergen-Paap rk Wald F-stat) evaluates the strength of the instrument; values greater than ten are considered strong (Stock & Yogo, 2005). The Endogeneity Test (Durbin-Wu-Hausman Chi2) determines whether the endogenous regressor should be instrumented; if the p-value exceeds 0.1, it indicates that that instrumenting may not be necessary. While the KP tests confirm that the instrument is strong across all six models, the Durbin-Wu-Hausman tests for the baseline and full GDP models suggest significant endogeneity concerns for $ECI^{software}$. Additionally, the coefficient for $ECI^{software}$ in the full model for Emissions becomes insignificant when additional complexity measures ($ECI^{trade}$, $ECI^{technology}$, $ECI^{research}$) are included.

## S9 ECI software, inequality, emission and the Kuznets curve

To test the hypotheses behind the Kuznets curve, the following tables show our main regressions on income inequality, emissions and ECI software using the quadratic term of GDP per capita. Our results are mixed in the context of income inequality, while when ECI software is included, none of the models indicate an inverted U-shaped relationship between emissions and GDP per capita.

**Table S9.1** Regressions on Gini coefficient including the quadratic term of GDP per capita

| | Gini coefficient | | | | | | | |
|---|---|---|---|---|---|---|---|---|
| | (1) | (2) | (3) | (4) | (5) | (6) | (7) | (8) |
| $ECI^{software}$ | -1.019*** | | | | -0.882** | -1.011** | -0.926*** | -0.864*** |
| | (0.351) | | | | (0.352) | (0.401) | (0.333) | (0.355) |
| $ECI^{trade}$ | | -0.689** | | | -0.514** | | | -0.286 |
| | | (0.274) | | | (0.253) | | | (0.255) |
| $ECI^{technology}$ | | | -0.221 | | | -0.019 | | 0.015 |
| | | | (0.245) | | | (0.268) | | (0.244) |
| $ECI^{research}$ | | | | 0.567*** | | | 0.525*** | 0.479** |
| | | | | (0.169) | | | (0.162) | (0.179) |
| GDP per capita (ln) | 7.673 | 8.249* | 7.668* | 12.779** | 8.264* | 7.694 | 12.627*** | 12.505*** |
| | (4.584) | (4.328) | (4.142) | (5.216) | (4.698) | (4.618) | (4.282) | (4.306) |
| GDP per capita$^2$ | -0.342 | -0.385* | -0.373* | -0.667** | -0.356 | -0.342 | -0.619*** | -0.602** |
| | (0.226) | (0.216) | (0.210) | (-0.270) | (0.229) | (0.227) | (0.221) | (0.222) |
| Population (ln) | 0.476*** | 0.253*** | 0.207** | 0.128* | 0.504*** | 0.477*** | 0.419*** | 0.438*** |
| | (0.134) | (0.087) | (0.091) | (0.072) | (0.132) | (0.132) | (0.104) | (0.105) |
| Natural resources (ln) | 0.180 | 0.206* | 0.276** | 0.285*** | 0.151 | 0.182 | 0.211** | 0.190 |
| | (0.107) | (0.103) | (0.109) | (0.074) | (0.106) | (0.113) | (0.079) | (0.094) |
| Observations | 48 | 48 | 48 | 48 | 48 | 48 | 48 | 48 |
| $R^2$ | 0.434 | 0.389 | 0.329 | 0.461 | 0.472 | 0.434 | 0.557 | 0.567 |

**Table S9.2** Regressions on emission including the quadratic term of GDP per capita

| | Emission per GDP (log) | | | | | | | |
|---|---|---|---|---|---|---|---|---|
| | (1) | (2) | (3) | (4) | (5) | (6) | (7) | (8) |
| $ECI^{software}$ | -0.113*** | | | | -0.116** | -0.103** | -0.079* | -0.072 |
| | (0.041) | | | | (0.043) | (0.047) | (0.045) | (0.051) |
| $ECI^{trade}$ | | -0.022 | | | 0.011 | | | 0.001 |
| | | (0.040) | | | (0.041) | | | (0.043) |
| $ECI^{technology}$ | | | -0.054 | | | -0.018 | | -0.014 |
| | | | (0.034) | | | (0.038) | | (0.040) |
| $ECI^{research}$ | | | | -0.065*** | | | -0.047** | -0.046* |
| | | | | (0.021) | | | (0.023) | (0.024) |
| GDP per capita (ln) | 0.282 | 0.306 | 0.362 | -0.133 | 0.268 | 0.310 | -0.014 | 0.011 |
| | (0.392) | (0.440) | (0.429) | (0.385) | (0.398) | (0.400) | (0.400) | (0.417) |
| GDP per capita$^2$ | -0.014 | -0.018 | -0.019 | 0.005 | -0.013 | -0.015 | 0.001 | 0.001 |
| | (0.020) | (0.023) | (0.022) | (0.020) | (0.020) | (0.020) | (0.020) | (0.021) |
| Population (ln) | 0.031* | -0.004 | 0.007 | -0.002 | 0.030 | 0.032* | 0.023 | 0.025 |
| | (0.018) | (0.014) | (0.016) | (0.013) | (0.018) | (0.018) | (0.018) | (0.018) |
| Natural resources (ln) | 0.053*** | 0.064*** | 0.065*** | 0.062*** | 0.055*** | 0.054*** | 0.053*** | 0.054*** |
| | (0.013) | (0.015) | (0.012) | (0.013) | (0.014) | (0.013) | (0.014) | (0.015) |
| Observations | 92 | 92 | 92 | 92 | 92 | 92 | 92 | 92 |
| $R^2$ | 0.555 | 0.510 | 0.525 | 0.557 | 0.555 | 0.557 | 0.576 | 0.577 |

## S10 Alternative entry and exit regression specifications

**Table S10.1** Logit regressions on countries gaining revealed comparative advantage (RCA >= 1) in software bundles (2020-2023). Standard errors are clustered at the country level. Significance codes: *p<0.1, **p<0.05, ***p<0.01

| | Entry | | | | | | |
|---|---|---|---|---|---|---|---|
| | (1) | (2) | (3) | (4) | (5) | (6) | (7) |
| Relatedness density | 2.754*** | 6.103*** | 5.594*** | 15.415*** | | 2.789*** | 5.727** |
| | (1.058) | (2.347) | (1.496) | (5.268) | | (1.005) | (2.541) |
| Ubiquity | | | | | -0.112 | -0.174*** | -0.194*** |
| | | | | | (0.180) | (0.165) | (0.202) |
| Country FE | No | Yes | No | Yes | No | No | Yes |
| Software bundle FE | No | No | Yes | Yes | No | No | No |
| Observations | 764 | 288 | 416 | 159 | 764 | 764 | 288 |
| Pseudo $R^2$ | 0.029 | 0.253 | 0.139 | 0.338 | 0.001 | 0.032 | 0.146 |
| BIC | 329 | 358 | 373 | 365 | 338 | 335 | 363 |

**Table S10.2** Logit regressions on countries losing revealed comparative advantage (RCA < 1) in software bundles (2020-2023). Standard errors are clustered at the country level. Significance codes: *p<0.1, **p<0.05, ***p<0.01

| | Exit | | | | | | |
|---|---|---|---|---|---|---|---|
| | (1) | (2) | (3) | (4) | (5) | (6) | (7) |
| Relatedness density | -3.505*** | -7.092*** | -3.727*** | -8.624*** | | -4.143*** | -6.230*** |
| | (0.580) | (1.646) | (0.675) | (2.866) | | (0.559) | (1.751) |
| Ubiquity | | | | | -0.126 | -0.476*** | -0.232 |
| | | | | | (0.133) | (0.157) | (0.198) |
| Country FE | No | Yes | No | Yes | No | No | Yes |
| Software bundle FE | No | No | Yes | Yes | No | No | No |
| Observations | 1544 | 778 | 1159 | 543 | 1544 | 1544 | 778 |
| Pseudo $R^2$ | 0.059 | 0.179 | 0.163 | 0.296 | 0.001 | 0.080 | 0.182 |
| BIC | 585 | 734 | 724 | 833 | 620 | 580 | 739 |

### S11 Robustness to RCA thresholds

Our main exercise follows the standard revealed comparative advantage threshold (RCA>=1) when binarizing the specialization matrix (Balland et al., 2022; Hidalgo, 2021). However, robustness checks were also performed using alternative thresholds, such as RCA>=0.75 and RCA>=1.25. Applying different thresholds results in minor changes (7% of country, software bundle combinations have different binary RCA values), which does not affect our main results.

**Table S11.1** Changes in the binarized Revealed Comparative Advantage (RCA) values for different thresholds

| Binary RCA (threshold = 1) | Binary RCA (threshold = 0.75) | Nr. country-software bundle pairs | Share |
|---|---|---|---|
| 0 | 0 | 5876 | 70% |
| 0 | 1 | 589 | 7% |
| 1 | 1 | 1914 | 23% |
| **Binary RCA (threshold = 1)** | **Binary RCA (threshold = 1.25)** | **Country-language pairs** | **Share** |
| 0 | 0 | 6465 | 77% |
| 1 | 0 | 596 | 7% |
| 1 | 1 | 1318 | 16% |

**Table S11.2** Correlation of software complexity values (ECI software) for different thresholds

| | ECI (RCA threshold=1) | ECI (RCA threshold=0.75) | ECI (RCA threshold=1.25) |
|---|---|---|---|
| ECI (RCA threshold=1) | 1 | 0.979 | 0.903 |
| ECI (RCA threshold=0.75) | 0.979 | 1 | 0.881 |
| ECI (RCA threshold=1.25) | 0.903 | 0.881 | 1 |

**Table S11.3** ECI software with different thresholds and GDP per capita (2020) in multidimensional settings

| | GDP per capita (log) | | | | | |
|---|---|---|---|---|---|---|
| | (1) | (2) | (3) | (4) | (5) | (6) |
| $ECI^{software}$ (threshold = 1.00) | 0.343*** | 0.125*** | | | | |
| | (0.025) | (0.044) | | | | |
| $ECI^{software}$ (threshold = 0.75) | | | 0.372*** | 0.151*** | | |
| | | | (0.027) | (0.051) | | |
| $ECI^{software}$ (threshold = 1.25) | | | | | 0.416*** | 0.055 |
| | | | | | (0.086) | (0.074) |
| $ECI^{trade}$ | | 0.190*** | | 0.182*** | | 0.213*** |
| | | (0.046) | | (0.046) | | (0.050) |
| $ECI^{technology}$ | | 0.063* | | 0.056 | | 0.090** |
| | | (0.035) | | (0.036) | | (0.036) |
| $ECI^{research}$ | | 0.022 | | 0.022 | | 0.042 |
| | | (0.026) | | (0.026) | | (0.026) |
| Population (ln) | -0.146*** | -0.122*** | -0.181*** | -0.128*** | -0.127*** | -0.096*** |
| | (0.017) | (0.016) | (0.018) | (0.020) | (0.019) | (0.016) |
| Natural resources (ln) | 0.015 | 0.028** | 0.017 | 0.029** | 0.005 | 0.023 |
| | (0.012) | (0.014) | (0.013) | (0.014) | (0.015) | (0.014) |
| Observations | 93 | 93 | 93 | 93 | 93 | 93 |
| $R^2$ | 0.648 | 0.764 | 0.671 | 0.770 | 0.531 | 0.748 |

**Table S11.4** ECI software with different thresholds and income inequality in multidimensional settings. ECI estimates are based on 2020 data, while the dependent variable is the average Gini coefficient between 2020 and 2022

| | Gini coefficient | | | | | |
|---|---|---|---|---|---|---|
| | (1) | (2) | (3) | (4) | (5) | (6) |
| $ECI^{software}$ (threshold = 1.00) | -1.038*** | -0.920** | | | | |
| | (0.353) | (0.381) | | | | |
| $ECI^{software}$ (threshold = 0.75) | | | -1.268*** | -1.200*** | | |
| | | | (0.397) | (0.383) | | |
| $ECI^{software}$ (threshold = 1.25) | | | | | -1.740*** | -1.572** |
| | | | | | (0.667) | (0.686) |
| $ECI^{trade}$ | | -0.359 | | -0.310 | | -0.339 |
| | | (0.293) | | (0.296) | | (0.303) |
| $ECI^{technology}$ | | 0.061 | | 0.103 | | -0.046 |
| | | (0.285) | | (0.263) | | (0.274) |
| $ECI^{research}$ | | 0.332** | | 0.356** | | 0.362** |
| | | (0.153) | | (0.152) | | (0.160) |
| GDP per capita (ln) | 0.905** | 0.759** | 0.972*** | 0.769* | 0.845** | 0.788** |
| | (0.350) | (0.343) | (0.342) | (0.384) | (0.349) | (0.343) |
| Population (ln) | 0.455*** | 0.422*** | 0.606*** | 0.567*** | 0.403*** | 0.399*** |
| | (0.129) | (0.113) | (0.168) | (0.137) | (0.133) | (0.113) |
| Natural resources (ln) | 0.250* | 0.279** | 0.290*** | 0.316*** | 0.258** | 0.299** |
| | (0.109) | (0.117) | (0.107) | (0.110) | (0.105) | (0.114) |
| Observations | 48 | 48 | 48 | 48 | 48 | 48 |
| $R^2$ | 0.409 | 0.499 | 0.436 | 0.530 | 0.389 | 0.493 |

**Table S11.5** ECI software with different thresholds and greenhouse gas emission intensity (2020) in multidimensional settings

| | Emission per GDP (log) | | | | | |
|---|---|---|---|---|---|---|
| | (1) | (2) | (3) | (4) | (5) | (6) |
| $ECI^{software}$ (threshold = 1.00) | -0.115** | -0.072 | | | | |
| | (0.041) | (0.050) | | | | |
| $ECI^{software}$ (threshold = 0.75) | | | -0.124*** | -0.081 | | |
| | | | (0.046) | (0.055) | | |
| $ECI^{software}$ (threshold = 1.25) | | | | | -0.191*** | -0.166*** |
| | | | | | (0.043) | (0.049) |
| $ECI^{trade}$ | | 0.001 | | 0.003 | | 0.020 |
| | | (0.042) | | (0.043) | | (0.042) |
| $ECI^{technology}$ | | -0.014 | | -0.013 | | -0.008 |
| | | (0.039) | | (0.040) | | (0.034) |
| $ECI^{research}$ | | -0.046** | | -0.047** | | -0.040** |
| | | (0.021) | | (0.021) | | (0.020) |
| GDP per capita (ln) | 0.011 | 0.019 | 0.016 | 0.022 | 0.010 | 0.014 |
| | (0.027) | (0.034) | (0.029) | (0.034) | (0.023) | (0.034) |
| Population (ln) | 0.031* | 0.025 | 0.043* | 0.033 | 0.032** | 0.030** |
| | (0.018) | (0.018) | (0.022) | (0.021) | (0.014) | (0.015) |
| Natural resources (ln) | 0.054*** | 0.054*** | 0.054*** | 0.054*** | 0.049*** | 0.050*** |
| | (0.013) | (0.015) | (0.013) | (0.014) | (0.012) | (0.013) |
| Observations | 92 | 92 | 92 | 92 | 92 | 92 |
| $R^2$ | 0.553 | 0.577 | 0.552 | 0.578 | 0.591 | 0.614 |

**Table S11.6** Entry models on countries gaining revealed comparative advantage (RCA) in programming languages (2020-2023) with different RCA thresholds

| | Entry (threshold 1.00) | | | Entry (threshold 0.75) | | | Entry (threshold 1.25) | | |
|---|---|---|---|---|---|---|---|---|---|
| | (1) | (2) | (3) | (4) | (5) | (6) | (7) | (8) | (9) |
| Relatedness density | 0.154** | 0.171** | 0.328** | 0.015 | 0.045 | 1.620** | 0.162*** | 0.162*** | 0.225*** |
| | (0.072) | (0.079) | (0.134) | (0.128) | (0.153) | (0.709) | (0.056) | (0.057) | (0.061) |
| Ubiquity | | -0.012 | -0.012 | | -0.017 | -0.079** | | -0.001 | -0.001 |
| | | (0.010) | (0.010) | | (0.027) | (0.036) | | (0.006) | (0.006) |
| Country FE | No | No | Yes | No | No | Yes | No | No | Yes |
| Observations | 764 | 764 | 764 | 304 | 304 | 304 | 1,356 | 1,356 | 1,356 |
| $R^2$ | 0.013 | 0.016 | 0.189 | 0.0001 | 0.003 | 0.433 | 0.012 | 0.012 | 0.089 |

**Table S11.7** Exit models on countries losing revealed comparative advantage (RCA) in programming languages (2020-2023) with different RCA thresholds

| | Exit (threshold 1.00) | | | Exit (threshold 0.75) | | | Exit (threshold 1.25) | | |
|---|---|---|---|---|---|---|---|---|---|
| | (1) | (2) | (3) | (4) | (5) | (6) | (7) | (8) | (9) |
| Relatedness density | -0.160*** | -0.223*** | -0.348*** | -0.051*** | -0.083*** | -0.222*** | -0.138*** | -0.138*** | -0.115*** |
| | (0.033) | (0.043) | (0.099) | (0.020) | (0.026) | (0.063) | (0.030) | (0.029) | (0.027) |
| Ubiquity | | -0.027*** | -0.018** | | -0.015** | -0.003 | | 0.007* | -0.013** |
| | | (0.008) | (0.009) | | (0.006) | (0.006) | | (0.004) | (0.006) |
| Country FE | No | No | Yes | No | No | Yes | No | No | Yes |
| Observations | 1,544 | 1,544 | 1,544 | 2,208 | 2,208 | 2,208 | 2,208 | 2,208 | 2,208 |
| $R^2$ | 0.023 | 0.035 | 0.187 | 0.005 | 0.011 | 0.126 | 0.025 | 0.027 | 0.135 |

### S12 Tobit regressions for ECI software and macroeconomic indicators

To assess the robustness of our findings, we implemented Tobit regressions with censoring thresholds tailored to each model while maintaining consistency across specifications. Tobit regressions on GDP per capita (log) are left-censored at 0. The Tobit model for the Gini coefficient uses a logit-transformed dependent variable: log(Gini / (1 - Gini)). Tobit regressions on Emission per GDP (log) account for the fact that emission per GDP values are always positive but very small; left censoring at -Inf ensures the model handles the lower bound correctly. The results confirm that our main conclusions remain unchanged, except for the models on income inequality, which perform less reliably due to the smaller sample of countries.

**Table S12.1** Tobit regressions for controlled correlations between ECI measures and macroeconomic indicators

| | **GDP per capita (log)** | | **Gini coefficient** | | **Emission per GDP (log)** | |
|---|---|---|---|---|---|---|
| | (1) | (2) | (3) | (4) | (5) | (6) |
| $ECI^{software}$ | 0.343*** | 0.125*** | -0.070 | -0.072 | -0.115*** | -0.072* |
| | (0.033) | (0.046) | (0.063) | (0.056) | (0.037) | (0.043) |
| $ECI^{trade}$ | | 0.190*** | | 0.034 | | 0.001 |
| | | (0.042) | | (0.074) | | (0.042) |
| $ECI^{technology}$ | | 0.063* | | 0.113 | | -0.014 |
| | | (0.037) | | (0.110) | | (0.034) |
| $ECI^{research}$ | | 0.022 | | 0.038 | | -0.046** |
| | | (0.023) | | (0.035) | | (0.021) |
| GDP per capita (ln) | | | 0.056 | -0.298 | 0.446*** | 0.454*** |
| | | | (0.057) | (0.391) | (0.034) | (0.041) |
| Population (ln) | -0.146*** | -0.112*** | 0.024 | -0.041 | 0.465*** | 0.459*** |
| | (0.018) | (0.016) | (0.024) | (0.102) | (0.018) | (0.018) |
| Natural resources (ln) | 0.015 | 0.028** | 0.023 | -0.003 | 0.054*** | 0.054*** |
| | (0.013) | (0.011) | (0.023) | (0.076) | (0.010) | (0.011) |
| Observations | 93 | 93 | 48 | 48 | 92 | 92 |
| Log Likelihood | 10.429 | 29.170 | -0.305 | 3.126 | 36.264 | 38.815 |

**S13 Regressions for ECI software and macroeconomic indicators based on identical samples**

In our regressions on different ECI values and macroeconomic outcomes, the number of observations differs due to data availability. The table below reports regressions based on a sample of 48 countries for which data are available for all three dependent variables in 2020. While our results for GDP per capita and income inequality were unchanged, the 48% decline in observations affected the models for emissions.

**Table S13.1 ECI software and macroeconomic indicators using identical samples**

| | GDP per capita (log) | | Gini coefficient | | Emission per GDP | |
|---|---|---|---|---|---|---|
| | (1) | (2) | (3) | (4) | (5) | (6) |
| $ECI^{software}$ | 0.302*** | 0.147*** | -1.038*** | -0.920** | 0.001 | -0.001 |
| | (0.041) | (0.047) | (0.353) | (0.381) | (0.060) | (0.059) |
| $ECI^{trade}$ | | 0.129*** | | -0.359 | | 0.012 |
| | | (0.041) | | (0.293) | | (0.051) |
| $ECI^{technology}$ | | 0.031 | | 0.061 | | -0.029 |
| | | (0.042) | | (0.285) | | (0.045) |
| $ECI^{research}$ | | 0.067** | | 0.332** | | -0.070** |
| | | (0.028) | | (0.153) | | (0.027) |
| GDP per capita (ln) | | | 0.905*** | 0.759*** | -0.076 | 0.006 |
| | | | (0.350) | (0.343) | (0.064) | (0.078) |
| Population (ln) | -0.110*** | -0.086*** | 0.455*** | 1.224*** | 0.017 | 0.029 |
| | (0.017) | (0.017) | (0.129) | (0.293) | (0.026) | (0.024) |
| Natural resources (ln) | -0.025 | -0.001 | 0.250** | 0.486* | 0.059** | 0.052** |
| | (0.018) | (0.016) | (0.109) | (0.268) | (0.022) | (0.024) |
| Observations | 48 | 48 | 48 | 48 | 48 | 48 |
| $R^2$ | 0.774 | 0.862 | 0.409 | 0.499 | 0.497 | 0.580 |

## S14 VIF values behind our main regressions

To be transparent about the potential multicollinearity underlying our models, we report variance inflation factor (VIF) values for all our OLS regressions on ECI software and macroeconomic indicators such as GDP per capita, Gini coefficient and Emission per GDP. The tables indicate no issues of multicollinearity.

**Table S14.1** VIF values for OLS regressions on GDP per capita

| | GDP per capita | | | | | | | |
|---|---|---|---|---|---|---|---|---|
| | (1) | (3) | (4) | (5) | (6) | (7) | (8) | (9) |
| $ECI^{software}$ | 1.677 | | | | 3.154 | 3.760 | 2.452 | 4.917 |
| $ECI^{trade}$ | | 1.567 | | | 2.949 | | | 4.292 |
| $ECI^{technology}$ | | | 1.201 | | | 2.692 | | 3.846 |
| $ECI^{research}$ | | | | 1.168 | | | 1.708 | 1.783 |
| Population (ln) | 1.281 | 1.032 | 1.079 | 1.024 | 1.406 | 1.326 | 1.381 | 1.545 |
| Natural resources (ln) | 1.556 | 1.604 | 1.199 | 1.194 | 1.685 | 1.591 | 1.556 | 1.901 |

**Table S14.2** VIF values for OLS regressions on Gini coefficient

| | Gini coefficient | | | | | | | |
|---|---|---|---|---|---|---|---|---|
| | (1) | (3) | (4) | (5) | (6) | (7) | (8) | (9) |
| $ECI^{software}$ | 4.641 | | | | 4.898 | 5.082 | 4.664 | 5.174 |
| $ECI^{trade}$ | | 3.207 | | | 3.384 | | | 4.352 |
| $ECI^{technology}$ | | | 2.378 | | | 2.604 | | 3.112 |
| $ECI^{research}$ | | | | 1.909 | | | 1.918 | 2.143 |
| GDP per capita (ln) | 4.434 | 3.827 | 3.667 | 2.472 | 5.571 | 5.271 | 5.249 | 7.227 |
| Population (ln) | 3.278 | 1.442 | 1.623 | 1.186 | 3.325 | 3.353 | 3.372 | 3.491 |
| Natural resources (ln) | 1.844 | 1.795 | 1.773 | 1.813 | 1.879 | 1.955 | 1.946 | 2.290 |

**Table S14.3** VIF values for OLS regressions on emission per GDP

| | Emission per GDP | | | | | | | |
|---|---|---|---|---|---|---|---|---|
| | (1) | (3) | (4) | (5) | (6) | (7) | (8) | (9) |
| $ECI^{software}$ | 3.575 | | | | 3.861 | 4.436 | 4.285 | 5.248 |
| $ECI^{trade}$ | | 4.007 | | | 4.328 | | | 5.444 |
| $ECI^{technology}$ | | | 2.650 | | | 3.288 | | 3.957 |
| $ECI^{research}$ | | | | 1.440 | | | 1.726 | 1.838 |
| GDP per capita (ln) | 2.796 | 3.265 | 2.847 | 1.575 | 4.035 | 3.417 | 2.797 | 4.234 |
| Population (ln) | 2.219 | 1.347 | 1.546 | 1.278 | 2.265 | 2.269 | 2.307 | 2.366 |
| Natural resources (ln) | 1.530 | 1.697 | 1.217 | 1.267 | 1.869 | 1.541 | 1.531 | 2.038 |

## S15 GDP growth regressions

Our empirical analysis is based on the recent GitHub Innovation Graph dataset, which is only available for a short period (2020-2023). This does not allow us to perform robust growth models or nuanced time-series regressions. The table below presents growth models for the period 2020-2023 using GDP values in current USD (GDP PPP is only available until 2022) in a similar fashion as (Hidalgo and Hausmann, 2009). The models do not perform as expected and do not confirm the otherwise well documented relationship between GDP growth and ECI trade (see model 3). This result is attributed to the short period available.

**Table S15.1** Regressions on GDP growth for the period of 2020-2023

| | GDP growth (log, 2020-2023) | | | | | | | | |
|---|---|---|---|---|---|---|---|---|---|
| | (1) | (2) | (3) | (4) | (5) | (6) | (7) | (8) | (9) |
| $ECI^{software}$ | | 0.007 | | | | 0.001 | -0.004 | 0.006 | -0.009 |
| | | (0.016) | | | | (0.015) | (0.017) | (0.016) | (0.016) |
| $ECI^{trade}$ | | | 0.019 | | | 0.019 | | | 0.015 |
| | | | (0.017) | | | (0.017) | | | (0.020) |
| $ECI^{technology}$ | | | | 0.018 | | | 0.019 | | 0.014 |
| | | | | (0.012) | | | (0.012) | | (0.014) |
| $ECI^{research}$ | | | | | 0.003 | | | 0.002 | 0.006 |
| | | | | | (0.010) | | | (0.010) | (0.011) |
| GDP (log) | -0.040*** | -0.035 | -0.049 | -0.052 | -0.031 | -0.050 | -0.050 | -0.036 | -0.062 |
| | (0.014) | (0.034) | (0.032) | (0.035) | (0.034) | (0.037) | (0.038) | (0.037) | (0.044) |
| Population (ln) | | -0.003 | 0.003 | 0.002 | -0.003 | 0.003 | 0.002 | -0.002 | 0.008 |
| | | (0.012) | (0.012) | (0.013) | (0.014) | (0.013) | (0.013) | (0.014) | (0.016) |
| Natural resources (ln) | | 0.010*** | 0.012*** | 0.010*** | 0.010*** | 0.012*** | 0.009** | 0.010*** | 0.011** |
| | | (0.004) | (0.004) | (0.004) | (0.004) | (0.004) | (0.004) | (0.004) | (0.005) |
| Observations | 92 | 92 | 92 | 92 | 92 | 92 | 92 | 92 | 92 |
| $R^2$ | 0.111 | 0.161 | 0.173 | 0.177 | 0.160 | 0.173 | 0.177 | 0.161 | 0.184 |

## S16 Correlation of $ECI_{software}$ values

**Table S16.1** Correlation of $ECI_{software}$ values along the available period (2020-2023)

| Year | **2020** | **2021** | **2022** | **2023** |
|---|---|---|---|---|
| 2020 | 1.00 | 0.98 | 0.95 | 0.92 |
| 2021 | 0.98 | 1.00 | 0.97 | 0.94 |
| 2022 | 0.95 | 0.97 | 1.00 | 0.97 |
| 2023 | 0.92 | 0.94 | 0.97 | 1.00 |